\documentclass[journal]{IEEEtran}
\usepackage{amssymb}
\usepackage[cmex10]{amsmath}
\usepackage{stfloats}
\usepackage{graphicx}
\usepackage{subfigure}
\usepackage{tabularx}
\usepackage{epsfig,epsf,color,balance,cite}
\usepackage{verbatim}
\usepackage{url}
\usepackage{bm}

\newtheorem{theorem}{Theorem}

\newtheorem{lemma}{Lemma}
\usepackage{algorithm}
\usepackage{algpseudocode}
\begin{document}

\title{Association and Load Optimization with User Priorities in Load-Coupled Heterogeneous Networks}

\author{
\IEEEauthorblockN{Zhaohui Yang,
                  Wei Xu, \IEEEmembership{Senior Member, IEEE},
                  Jianfeng Shi,
                  Hao Xu, and
                  Ming Chen
                  }
                  \vspace{-2em}
\thanks{
This paper was presented at the Globecom 2016 in Washington, DC, USA \cite{yang2016}. 
}
\thanks{Z. Yang, W. Xu, J. Shi, H. Xu and M. Chen are with the National Mobile Communications Research
Laboratory, Southeast University, Nanjing 210096, China.  (Email: yangzhaohui@seu.edu.cn; wxu@seu.edu.cn; shijianfeng@seu.edu.cn; xuhao2013@seu.edu.cn; chenming@seu.edu.cn).}
}

\maketitle

\begin{abstract}
In this paper, we consider the network utility maximization problem with various user priorities via jointly optimizing user association, load distribution and power control in a load-coupled heterogeneous network. In order to tackle the nonconvexity of the problem, we first analyze the problem by obtaining the optimal resource allocation strategy in closed form and characterizing the optimal base station load distribution pattern. Both observations are shown essential in simplifying the original problem and making it possible to transform the nonconvex load distribution and power control problem into convex reformulation via exponential variable transformation. An iterative algorithm with low complexity is accordingly presented to obtain a suboptimal solution to the joint optimization problem. Simulation results show that the proposed algorithm achieves better performance than conventional approaches.
\end{abstract}

\IEEEpeerreviewmaketitle

\section{Introduction}
To meet the growing demand for high data rate transmission and seamless coverage in wireless communications, heterogeneous deployment is introduced in the 5G network \cite{6824752,andrews2012femtocells,dhillon2012modeling,wang2014cellular,demestichas20135g}.
In heterogeneous networks (HetNets), small base stations (BSs) are deployed to offload the traffic for users in high user density area.
The small BSs usually share the same frequency band as macro BSs to improve overall spectrum efficiency of the entire network.

User association has been one of the main challenges in the deployment of HetNets.
Conventionally, a typical scheme is based on the Max-SINR (signal-to-interference-plus-noise ratio) rule, i.e., each user is associated with the BS that provides the highest SINR.
Although the Max-SINR rule is straightforward, it simply prevents the incorporation of multiple network requirements such as load balancing and minimal rate constraints.
To overcome this shortcoming,
the earlier works \cite{wang2010sinr,smolyar2009unified} focused on the total transmit power minimization problem with individual user rate constraints in terms of minimal SINR requirements.
On the other hand, there are a number of works aiming at the overall throughput maximization \cite{corroy2012dynamic,yinghong2015cell,boostanimehr2015joint,boostanimehr2015unified, zhou2015distributed}.
A more general network utility maximization problem was proposed in \cite{ye2013user}, which utilized the logarithmic utility function and proved that equal resource allocation is in fact optimal.
Then, the logarithmic utility maximization problem for joint user association and power control was studied in \cite{shen2014distributed}.
Recently, \cite{chenuser} further studied the user association problem with various user priorities.

Since HetNets are ususlly based on orthogonal frequency-division multiplexing (OFDM), both resource allocation and power control are essential for inter-cell interference suppression.
In HetNets, the load (average utilization level of the time-frequency resource blocks) conditions in macro BSs and small BSs are in most case coupled.
A realistic load-coupled model was proposed in \cite{Siomina2009Math,Majewski2010Conservative,Siomin2012Analysis,7562205}, which took into account the effect of the load conditions in terms of inter-cell interference.
In \cite{fehske2012aggregation}, it was shown that this load-coupled model well models a multi-cell network.
There are many works considering the load-coupled model in HetNets \cite{siomina2014constrained,siomina2014optimal,chen2015learning,chen2015energy}.
In \cite{siomina2014constrained}, a utility maximization framework for data offloading in load-coupled HetNets was proposed.
The problem of setting cell load levels for maximizing the overall system utility was investigated in \cite{siomina2014optimal}.
The above existing works \cite{siomina2014constrained,siomina2014optimal,chen2015learning,chen2015energy} all assumed fixed user association in the load-coupled HetNet even though proper user association plays a critical role in achieving enhanced network performance especially in HetNets.
Recently,
the sum load minimization and maximum load minimization were studied in \cite{DBLPYouY16} with user association in load-coupled HetNets.
However, \cite{DBLPYouY16} assumed fixed transmit power of BSs even though power control strategy is critical in HetNets for intra/inter-cell interference control.

In this paper, we consider joint user association, load distribution and power control for a load-coupled HetNet, where
a logarithmic utility objective is maximized for users having different priorities.
Since the logarithmic utility maximization problem with nonlinear equalities and discrete constraints is nonconvex, it is in general difficult to achieve the optimal solution.
We decompose the logarithmic utility maximization problem into three stages to obtain a suboptimal solution.
We first find the optimal user association with fixed load distribution and power control.
Then, we obtain the optimal load distribution and power control assuming equal power among users within the same cell.
On top of it, the inner-cell power allocation is optimized to further exploit multiuser diversity.

The joint problem of user association and power control with unequal user priorities can partly be treated by using the existing methodologies in \cite{ye2013user,shen2014distributed,chenuser},
i.e., the idea is to iteratively conduct the user association method with fixed power, and then a power control method under fixed user association.
All these methods, however, only partly solve the problem even without guarantee in obtaining local optimum.
This is fundamentally due to the nonconvexity of the joint optimization problem, which is challenging to deal with as evidenced in \cite{ye2013user,shen2014distributed,chenuser}.
Apparently, even the subproblems in existing methods do not embrace the convexity property, thus they can not be exploited to solve the problem in a better way, e.g., efficiently and optimally.
From this perspective, we transform the nonconvex power control subproblem in \cite{shen2014distributed} into an equivalent convex problem, which can be optimally solved.
Moreover, we propose a method to solve the user association subproblem optimally with polynomial complexity, while it is solved with polynomial complexity in \cite{chenuser} suboptimally.
In addition to user association and power control, time-frequency resource allocation is another technique to manage inter-cell interference.
The joint design of user association, load distribution and power control can achieve better performance.
Since the user association variable, load variable and power variable are coupled in both the utility objective function and the average SINR formulation, the joint optimization problem becomes even more complicated.

In detail, the main contributions in this paper are summarized as follows:
\begin{enumerate}
  \item We formulate a logarithmic utility maximization problem for
  a load-coupled HetNet, and show that the optimal resource allocation for users is proportional to user priorities.
  With this finding, we devise a low-complexity distributed algorithm via dual decomposition to obtain the optimal user association with fixed load distribution and power control.
  \item It is revealed that the optimal BS load distribution is strictly binary, i.e.,
   the BS is either fully loaded or shut down, for maximizing the network utility function.
   The optimal load distribution helps us transform the original nonconvex power control problem into a convex one by applying some variable transformations.
   \item The logarithmic utility can be further enhanced by using power control within each cell.
   With inner-cell unequal power allocation, the performance of both low-rate users and high-rate users can be accordingly improved.
\end{enumerate}

This paper is organized as follows.
In Section $\text{\uppercase\expandafter{\romannumeral2}}$, we introduce the system model and provide the sum utility maximization problem formulation.
Section $\text{\uppercase\expandafter{\romannumeral3}}$ provides
the optimal conditions of sum utility maximization, and proposes an iterative user association, load distribution and power control algorithm.
In Section $\text{\uppercase\expandafter{\romannumeral4}}$, we provide the sum utility maximization problem with inner-cell unequal power allocation and propose a dual method to obtain the optimal resource allocation and power control.
Numerical results are displayed in Section $\text{\uppercase\expandafter{\romannumeral5}}$
and conclusions are finally drawn in Section $\text{\uppercase\expandafter{\romannumeral6}}$.

\section{System Model and Problem Formulation}

\subsection{System Model}

Consider a downlink HetNet with $I$  BSs and $J$ users, as illustrated in Fig.~1.
Let $\mathcal I=\{1, 2, \cdots, I\}$ and $\mathcal J=\{1, 2, \cdots, J\}$ be the sets of all BSs and users, respectively.
Denote $x_{ij}$ as the association for BS $i$ and user $j$, i.e., $x_{ij}=1$ when user $j$ is associated with BS $i$; otherwise, $x_{ij}=0$.
Assuming that each user is associated with only one BS, it gives
\begin{equation}\label{SysEq1}
\sum_{i\in\mathcal I}x_{ij} = 1, \quad \forall j \in \mathcal J.
\end{equation}

\begin{figure}
  \centering
  \includegraphics[width=2.8in]{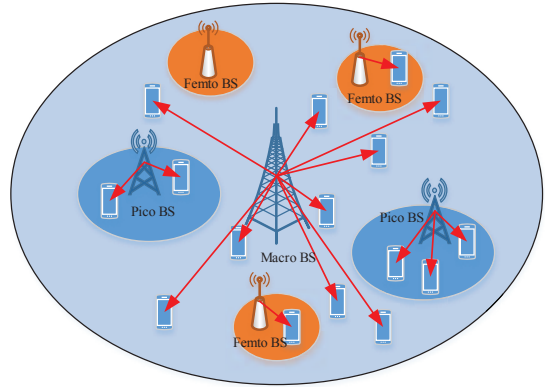}\\
  \caption{System model.}
\end{figure}

Let $y_{ij}$ denote the fraction of time-frequency resource blocks allocated to user $j$ by BS $i$.
Assume that all BSs in different tiers share the same number of time-frequency resource blocks, of which the total number is $K$.
A number of $K y_{ij}$ time-frequency resource blocks are allocated to user $j$.
Obviously, $y_{ij}>0$ if and only if user $j$ is associated with BS $i$, which implies that $y_{ij}\leq x_{ij}$.
Assume that multiple users associated to the same BS are allocated with orthogonal time-frequency resource blocks.
Denote the load of BS $i$ by $d_i$, which is defined to be the proportion of time-frequency resource blocks consumed by BS $i$ due to serving all users associated with it \cite{Siomin2012Analysis}.
The load, $d_i$, of BS $i$ can be evaluated by summing the fractions of time-frequency resource blocks occupied by users associated with the BS, i.e.,
\begin{equation}\label{SysEq2}
d_{i}=\sum_{j\in\mathcal J}y_{ij}\leq 1, \quad\forall i \in \mathcal I.
\end{equation}

The load-coupled model is helpful in characterizing the inter-cell interference especially for a multi-cell network
with OFDM \cite{Siomin2012Analysis,fehske2012aggregation}.
Denote $p_{i}$ as the transmit power of BS $i$ on each resource block.
It is assumed that the transmit power of BS $i$ is the same for all users associated with BS $i$.
The case that users in the same cell are allocated with unequal transmit power is discussed in Section $\text{\uppercase\expandafter{\romannumeral4}}$.
Given that the resource blocks are allocated to users randomly and we consider the long-term average interference from other BSs,
the average SINR of user $j$ associated with BS $i$ can be expressed as \cite{Siomin2012Analysis}
\begin{equation}\label{SysEq3}
\eta_{ij} \protect=\frac{p_{i} g_{ij}} {\sum_{k \in {\cal{I}}\setminus \{i\}} d_k
 p_{k} g_{kj} +\sigma ^2},
\end{equation}
where $g_{ij}$ is the channel gain from BS $i$ to user $j$ and $\sigma^2$ represents the power of the additive white Gaussian noise (AWGN) on each resource block.
Since BS $k$ ($k\neq i$) with high load level results in that BS $k$ utilizes the same time-frequency resource blocks as BS $i$ with high probability,
$d_k$ can be interpreted as the probability of receiving interference from BS $k$ across all the time-frequency resource blocks \cite{Siomin2012Analysis}.
Thus, the term $d_k p_k g_{kj}\in [0, p_k g_{kj}]$ is interpreted as the average interference taken over time and frequency for all transmissions.
Formula (\ref{SysEq3}) with averaged interference power evaluated by load variables has been shown to give a good approximation for a multi-cell network especially at high data arrival rates \cite{fehske2012aggregation}.
Note that this formulation has also been used in a number of applications like in \cite{siomina2014constrained,siomina2014optimal,chen2015learning,chen2015energy,DBLPYouY16}
because of its good structure with high accuracy characterizing inter-cell interference.
The achievable rate, $r_{ij}$, of user $j$ associated with BS $i$ can be formulated as
\begin{equation}\label{SysEq4}
r_{ij}=K B y_{ij}\log_2(1+\eta_{ij}),
\end{equation}
where $B$ is the bandwidth of each time-frequency resource block.
Note that $\eta_{ij}$ defined in (\ref{SysEq3}) is an average SINR, and the achievable rate $r_{ij}$ in (\ref{SysEq4}) can be regarded as a lower bound of the average achievable rate due to the convexity of function $\log_2\left( 1+ \frac 1 x\right)$.
Owing to the fact that user $j$ can be associated with any single BS, the effective rate of user $j$ is written as
\begin{equation}\label{SysEq4}
r_{j}=\sum_{i\in\mathcal {I}}r_{ij}=KB\sum_{i\in\mathcal I}y_{ij}\log_2(1+\eta_{ij}).
\end{equation}

\subsection{Problem Formulation}

We aim at network utility maximization via adjusting user association, load distribution and power control with different user priorities, where the priority of user $j$ is denoted by $\omega_j$.
Here, $\omega_j$ is a positive constant, which reflects the physical feature of user $j$.
Now it is ready to formulate the utility maximization problem as
\begin{subequations}\label{max1_1}
\begin{align}
\mathop{\max}_{\pmb{x}, \pmb{y}, \pmb{p}}\quad\!\!
&\sum_{j \in {\cal{J}}}\omega_j U\left(KB\sum_{i\in {\cal{I}}}y_{ij}\log_2(1+\eta_{ij})\right)
\\
\textrm{s.t.}\quad\!\!
&\eta_{ij}=\frac{p_{i} g_{ij}} {\sum_{k \in {\cal{I}}\setminus \{i\}} d_k p_{k} g_{kj} +\sigma ^2}, \quad\forall i  \in {\cal{I}}, j \in \mathcal J\\
&\sum_{i\in\mathcal I}x_{ij} = 1, \quad\forall j \in \mathcal J\\
&d_i= \sum_{j\in {\cal{J}}}y_{ij}, \quad\forall i  \in {\cal{I}}\\
&0 \leq p_{i} \leq P_i, \quad\forall i \in \mathcal I\\
&0 \leq d_i \leq 1, \quad\forall i \in \mathcal I \\
&0\leq y_{ij}\leq x_{ij}, x_{ij}\in\{0, 1\}, \quad\forall i  \in {\cal{I}}, j \in \mathcal J,
\end{align}
\end{subequations}
where $\pmb x=[x_{11}, \cdots, x_{1J}, \cdots, x_{IJ}]^T$ is the user association vector,
$\pmb y=[y_{11}, \cdots,$ $y_{1J}, \cdots,$ $y_{IJ}]^T$ is the resource allocation vector,
$\pmb p=[p_1, \cdots, p_I]^T$ is the power control vector,
$U(\cdot)$ is the utility function,
and
$P_i$ is the maximal transmit power of BS $i$ on each resource block.

In order to realize proportional fairness, we choose the logarithmic utility function $U(x)=\log_2(x), \forall j \in \mathcal J$ as in \cite{ye2013user,shen2014distributed,chenuser}.
The logarithm function is concave which has diminishing returns.
This property encourages load balancing  and  fairness among users.
Since problem (\ref{max1_1}) with nonlinear equalities (\ref{max1_1}b) and discrete constraints (\ref{max1_1}g) is nonconvex, it is in general difficult to obtain the global optimum.

\section{Joint User Association, Load Distribution and Power Control}
In this section, we first provide the optimal conditions for the resource allocation and the load of each BS.
Then, a joint optimization algorithm is proposed with iterative mechanism.
The analysis of complexity is also provided for comparison.
\subsection{Optimal Conditions for the Resource Allocation and the Load Distribution }
In order to facilitate the solution to problem (\ref{max1_1}), we here first present some interesting observations in the following Theorem~1 on the optimal condition of the resource allocation and Theorem~2 on the optimal strategy for BS loading.
Note that both two theorems help us simplify the procedure of solving the problem in (\ref{max1_1}) without loss of optimality.

\begin{theorem}
Let the optimal solution to problem (\ref{max1_1}) be ($\pmb x^*, \pmb y^*, \pmb p^*$).
The optimal resource allocation vector satisfies
\begin{equation}\label{eqresour}
y_{ij}^*=
\frac{\omega_j x_{ij}^*d_i^*}{\sum_{l\in \mathcal J}\omega_l x_{il}^*}, \quad \forall i \in \mathcal I, j \in \mathcal J_i,
\end{equation}
where $d_i^*=\sum_{j \in \mathcal J} y_{ij}^*$, and we define $\frac{\omega_j x_{ij}^*d_i^*}{\sum_{l\in \mathcal J}\omega_l x_{il}^*}=0$  for the case $d_i^*=0$, $x_{ij}^*=0$, $\forall j \in \mathcal J$.
\end{theorem}
\itshape \textbf{Proof:}  \upshape Please refer to Appendix A. \hfill $\Box$

Theorem 1 states that the optimal resource
allocation is proportional to the user priority, while it is independent of the SINR distribution.
It is obvious that user with higher priority intends to be allocated with a larger fraction of resource blocks.
For the special case with equal priorities, i.e., $\omega_1=\cdots=\omega_J$, we can observe that the optimal resource allocation (\ref{eqresour}) becomes $y_{ij}^*=
\frac{x_{ij}^*d_i^*}{\sum_{l\in \mathcal J} x_{il}^*}$, which means that the
optimal allocation is uniform for users served by that BS.
This observation agrees with the previous conclusion in \cite[Theorem~1]{ye2013user} as a special case when all the priorities are the same.

\begin{theorem}
For problem (\ref{max1_1}), the optimal load for a BS is always binary, i.e., $d_i^*=\sum_{j\in\mathcal J} y_{ij}^*\in\{0,1\}, \forall i \in \mathcal I$.
It implies that the optimal load distribution in the network is that
each BS operates best at either full load, i.e., $d_i^*=1$, or zero load, i.e., $d_i^*=0$.
\end{theorem}

\itshape \textbf{Proof:}  \upshape Please refer to Appendix B. \hfill $\Box$

From the network interference control perspective, each BS operating in the binary on-off status can definitely reduce the number of active BSs within the network, and hence inter-cell interference in the network can be somewhat suppressed.
However, from the user association perspective, each user is likely to be associated with its nearest BS which definitely enjoys the best wireless channel gains.
Therefore, it is interesting, but not obvious, to see that the best BS load distribution within the networks is strictly restricted to be binary for maximizing the entire network utility.
Theorem 2 implies that the resource of a BS should be either fully used or the BS is shut down in order to maximize the network utility.
This conclusion can be instructive for actual operation of a HetNet.
Moreover, this observation fortunately obeys the rules for green networks where the smallest number of activated BSs can meanwhile save power consumption of the entire network.

Now we look at the original problem in (\ref{max1_1}) which is
 combinational due to the binary variable $x_{ij}$.
Solving a combinational problem is usually impossible even for a modest-sized cellular network \cite{Papadimitriou1982Combinatorial}.
We here temporarily adopt the fractional user association relaxation, where association variable $x_{ij}$ can take on any real value in [0,1].
It is important to note that the relaxation fortunately does not cause any loss of optimality to the final solution to the original problem in (\ref{max1_1}).
We will later show that the optimal solution to $x_{ij}$ must be either 1 or 0 even though the feasible region of $x_{ij}$ is relaxed to be continuous.
Given the optimal resource allocation in (\ref{eqresour}), the relaxed problem (\ref{max1_1}) can be formulated as
\begin{subequations}\label{max1_2}
\begin{align}
\mathop{\max}_{\pmb{x}, \pmb d, \pmb{p}} \:\:
&
\sum_{i \in \mathcal I}\sum_{j \in {\cal{J}}}\omega_j x_{ij}\log_2\left(\frac{KB \omega_j d_i \log_2(1+\eta_{ij}) }{\sum_{l\in \mathcal J}\omega_l x_{il}}\right)
\\
\textrm{s.t.} \quad\!\!
&\eta_{ij}=\frac{p_{i} g_{ij}} {\sum_{k \in {\cal{I}}\setminus \{i\}} d_k p_{k} g_{kj} +\sigma ^2}, \quad\forall i  \in {\cal{I}}, j \in \mathcal J\\
&\sum_{i\in\mathcal I}x_{ij} = 1, \quad\forall j \in \mathcal J\\
&0 \leq d_i \leq 1, \quad\forall i \in \mathcal I\\
&0 \leq p_{i} \leq P_i, \quad\forall i \in \mathcal I\\
& x_{ij} \geq 0, \quad\forall i  \in {\cal{I}}, \forall j \in \mathcal J,
\end{align}
\end{subequations}
where $\pmb d =[d_1, \cdots, d_I]^T$ is the load distribution vector.
From Theorem 2, the loads of some BSs can be zeros.
For the case with $d_i=0$ and $x_{ij}=0$, $\forall j\in \mathcal J$, we define
$x_{ij}\log_2\left(\frac{B \omega_j d_i \log_2(1+\eta_{ij})}{\sum_{l\in \mathcal J}\omega_l x_{il}}\right)=0$, $\forall j \in \mathcal J$.
The equivalence  of (\ref{max1_1}a) and (\ref{max1_2}a) follows from (\ref{AeqObj6a}) in Appendix B.

Note that problem (\ref{max1_2}) is still nonconvex, obtaining the globally optimal solution is a difficult task.
Instead of aiming at the global optimality, we present an iterative algorithm for solving the nonconvex problem.

\subsection{User Association with Fixed Load Distribution and Power Control}
Constraints (\ref{max1_2}b), (\ref{max1_2}d) and (\ref{max1_2}e) correspond only to variables $\pmb d$ and $\pmb p$.
Considering fixed $\pmb d$ in Theorem 2 and assuming fixed $\pmb p$, problem (\ref{max1_2}) becomes
\begin{subequations}\label{max1_2_21}
\begin{align}
\mathop{\max}_{\pmb{x}} \:\:
&
\sum_{i \in \mathcal I}\sum_{j \in {\cal{J}}}\omega_j x_{ij}\log_2\left(\frac{KB\omega_j d_i  \log_2(1+\eta_{ij})}{\sum_{l\in \mathcal J}\omega_l x_{il}}\right)
\\
\textrm{s.t.} \quad\!\!
&\sum_{i\in\mathcal I}x_{ij} = 1, \quad\forall j \in \mathcal J\\
& x_{ij} \geq 0, \quad\forall i  \in {\cal{I}}, \forall j \in \mathcal J.
\end{align}
\end{subequations}

\begin{theorem}
Even though problem (\ref{max1_2_21}) with discrete constraints $x_{ij}\in\{0, 1\}, \forall i \in \mathcal I, j \in \mathcal J,$ is nonconvex,
the optimal solution to problem (\ref{max1_2_21}) can be effectively solved via its dual problem, while satisfying the discrete constraints $x_{ij}\in\{0, 1\}, \forall i \in \mathcal I, j \in \mathcal J$.
\end{theorem}

\itshape \textbf{Proof:}  \upshape
Let $c_{ij}\triangleq\omega_j \log_2({KB\omega_j d_i  \log_2(1+\eta_{ij})})$ and denote $N_i\triangleq\sum_{j\in \mathcal J}\omega_j x_{ij}$.
We rewrite (\ref{max1_2_21}) in the following equivalent form
\begin{subequations}\label{max1_2_22}
\begin{align}
\mathop{\max}_{\pmb{x}, \pmb N} \:\:
&
\sum_{i \in \mathcal I}\sum_{j \in {\cal{J}}}c_{ij} x_{ij} -\sum_{i \in \mathcal I}N_i \log_2(N_i)
\\
\textrm{s.t.} \quad\!\!
&\sum_{i\in\mathcal I}x_{ij} = 1, \quad \forall j \in \mathcal J\\
&N_i=\sum_{j\in \mathcal J}\omega_j x_{ij}, \quad\forall i \in \mathcal I\\
&x_{ij}\geq 0, N_i\geq 0, \quad \forall i  \in {\cal{I}}, j \in \mathcal J,
\end{align}
\end{subequations}
where $\pmb N=[N_1, \cdots, N_I]^T$.
Since
\begin{equation}
-\frac{\partial^2 N_i \log_2(N_i)}
{\partial N_i^2}
=-\frac{1}{(\ln2)N_i}\leq0,
\end{equation}
 (\ref{max1_2_22}a) is concave.
Considering that constraints (\ref{max1_2_22}b)-(\ref{max1_2_22}d) are all linear, problem (\ref{max1_2_22}) is convex.

Denoting $\pmb \mu=[\mu_1, \cdots, \mu_I]^T$ as the Lagrange multiplier vector associated with constraints (\ref{max1_2_22}c), we obtain the dual problem of (\ref{max1_2_22}) as
\begin{equation}\label{max1_7}
\mathop{\min}_{\pmb{\mu}} \quad
D(\pmb\mu)=f_{\pmb x}(\pmb \mu)+g_{\pmb N}(\pmb \mu),
\end{equation}
where
\begin{equation}\label{max1_8}
f_{\pmb x}(\pmb \mu)=\left\{ \begin{array}{ll}
\!\!\!\mathop{\max}\limits_{\pmb{x}}
&\!\!
\sum\limits_{i \in \mathcal I}\sum\limits_{j \in {\cal{J}}}(c_{ij}-\omega_j \mu_i)x_{ij}
\\
\textrm{s.t.}
&\!\!\sum\limits_{i\in\mathcal I}x_{ij} = 1, \quad\forall j \in \mathcal J\\
&\!\! x_{ij}\geq 0,\quad \forall i  \in {\cal{I}}, j \in \mathcal J,
\end{array} \right.
\end{equation}
and
\begin{equation}\label{max1_9}
g_{\pmb N}(\pmb \mu)=\left\{ \begin{array}{ll}
\!\!\!\mathop{\max}\limits_{\pmb{N}}
&\!\!
\sum\limits_{i \in \mathcal I}N_i(\mu_i-\log_2(N_i))
\\
\textrm{s.t.}
&\!\!  N_{i}\geq 0, \quad\forall i  \in {\cal{I}}.
\end{array} \right.
\end{equation}

The constraints in convex problem (\ref{max1_2_22}) are all linear, and
thus the Slater condition holds \cite{boyd2004convex}.
Therefore,
the primal problem (\ref{max1_2_22}) can be equivalently solved by its dual problem in (\ref{max1_7}) with zero dual gap, i.e., the optimal value of (\ref{max1_2_22}) and (\ref{max1_7}) is the same.

Because both objective and constraints can be decoupled,
these two sub-problems (\ref{max1_8}) and (\ref{max1_9}) can be solved in a distributed manner.
Each user measures $c_{ij}$ by using the pilot signals and receives the value of $\mu_i$ broadcast by each BS.
By solving linear problem (\ref{max1_8}), the optimal user association for each user is directly given as
\begin{equation}\label{sol1}
x_{ij}(t+1)=\left\{ \begin{array}{ll}
\!\!1, &\text{if}\; i =\arg\max\limits_{k\in\mathcal I}(c_{kj}-\omega_j \mu_k(t))\\
\!\!0, &\text{otherwise},
\end{array} \right.
\end{equation}
where $t$ is iteration number.

Each BS updates the new value of $N_i$ and $\mu_i$ in two steps.
In the first step, $N_i$ is updated by
\begin{equation}\label{sol2}
N_{i}(t+1)=\text{e}^{(\ln2)\mu_i(t)-1},
\end{equation}
which is the solution to convex problem (\ref{max1_9}).
In the second step,
to solve the dual optimization problem (\ref{max1_7}),
we use the gradient method to update Lagrange multiplier $\mu_i$ according to
\begin{equation}\label{sol3}
 \mu_i{(t\!+\!1)}\!=\!\mu_i{(t)}\!-\!
\theta(t)\! \!\left(\!N_i(t+1)\!-\!\sum_{j\in \mathcal J}\omega_j x_{ij}{(t+1)}\!\right),
\end{equation}
where $\theta(t)>0$ is a dynamically chosen stepsize sequence.
We can adopt the typical self-adaptive scheme of \cite{bertsekas2009convex} to chose the dynamic stepsize.

In summary, by iteratively updating primary variables and dual variables, the dual gradient projection (DGP) algorithm yields the optimal solution to the primal user association problem in (\ref{max1_2_21}).
According to (\ref{sol1}) in each iteration, the optimal solution to $x_{ij}$ in problem (\ref{max1_2_21}) automatically satisfies the discrete constraints $x_{ij}\in\{0, 1\}, \forall i \in \mathcal I, j \in \mathcal J$.
\hfill $\Box$

\subsection{Load Distribution and Power Control with Fixed User Association}
\label{loadpowerFixedAssociation}
Since constraints (\ref{max1_2}c) and (\ref{max1_2}f) are only determined by user association $\pmb x$, the load distribution and power control problem (\ref{max1_2}) with fixed user association $\pmb x$ can be expressed as
\vspace{-0.5em}
\begin{subequations}\label{max1_3}\vspace{-0.5em}
\begin{align}
\mathop{\max}_{\pmb d, \pmb{p}} \;
&
\sum_{i \in \mathcal I}\sum_{j \in {\cal{J}}}\omega_j x_{ij}\log_2\left(d_i \log_2(1+\eta_{ij}) \right)+C
\\
\textrm{s.t.} \;\;
&\eta_{ij}= \frac{p_{i} g_{ij}} {\sum_{k \in {\cal{I}}\setminus \{i\}}d_k p_{k} g_{kj} +\sigma ^2}, \quad\forall i  \in {\cal{I}}, j \in \mathcal J\\
&0\leq d_i \leq 1, \quad\forall i \in \mathcal I \\
&0 \leq p_{i} \leq P_i, \quad \forall i \in \mathcal I,
\end{align}
\end{subequations}
where $C$ defined in (\ref{Aeqmax1_3Constant}) in Appendix B is a constant.

According to Lemma 1 in Appendix B, a BS should operate at full load if there exists at least one user associated to this BS, otherwise the BS shuts down, or equivalently with zero load.
This is reasonable and straightforward, because it is always energy saving to shut off a BS if there is no preferred user associated with this BS.
Since the optimal $\pmb d$ can be obtained from Lemma~1, we only need to obtain the optimal $\pmb p$ of problem (\ref{max1_3}).
In the following, we obtain the optimal solution to problem (\ref{max1_3}) by transforming it into an equivalent convex problem.

Obviously, if $d_i^*=0$, we obtain $p_{i}=0$ and $x_{ij}=0$, $\forall j \in \mathcal J$.
Thus, we only need to solve the power control problem for BSs with full load.
Denote $\mathcal A$ as the set of BSs with full load, i.e., $\mathcal A=\{i\in\mathcal I|d^*_i=1\}$, which has been directly determined through the user association solution.
Let $\mathcal J_i=\{j\in\mathcal J|x_{ij}=1\}$ denote the set of users associated with BS $i$.
Owing to the fact that each user is associated with one BS, there exists only one $i\in\mathcal A$ such that $x_{ij}=1$ for any user $j\in\mathcal J$.
Accordingly, for notational convenience, we can use $\eta_j$ to replace $\eta_{ij}$, $\forall j\in\mathcal J_i$, without loss of generality.
Treating $\eta_{j}$ as new variable, problem (\ref{max1_3}) with the optimal load distribution is
equivalent to the following problem:
\begin{subequations}\label{max1_3_2}
\begin{align}
\!\!\!\!\!\!
\mathop{\max}_{\pmb{p}, \pmb{\eta}} \;
&
\sum_{i\in\mathcal A}\sum_{j \in {\cal{J}}_i}\omega_j \ln\left( \ln(1+\eta_{j}) \right)
\\
\textrm{s.t.} \;\;
&0\!\leq\!\eta_{j}\!\leq\! \frac{p_{i} g_{ij}} {\sum_{k \in {\cal{A}}\setminus \{i\}}  p_{k} g_{kj} \!+\!\sigma ^2}, \!\!\quad \forall i\in\mathcal A, j \in \mathcal J_i\\
&0 \leq p_{i} \leq P_i, \quad \forall i \in \mathcal A,
\end{align}
\end{subequations}
where $\pmb \eta=[\eta_{1}, \cdots, \eta_{J}]^T$.

Obviously, problem (\ref{max1_3_2}) is nonconvex due to constraints (\ref{max1_3_2}b).
To address the difficulty, we introduce some exponential variable transformations, which have two advantages.
The first advantage is that the constraints $\eta_{j}\geq 0$ and $p_i \geq 0$ can be implicitly removed.
The other advantage is that nonconvex constraints (\ref{max1_3_2}b) can be transformed into convex constraints and objective function (\ref{max1_3_2}a) remains concave after the transformations.
Due to the above two advantages, the original nonconvex problem (\ref{max1_3_2}) can be transformed into a convex problem through the following exponential variable transformation.

Letting $\eta_{j}=\text{e}^{u_{j}}$ and  $p_{i}=\text{e}^{v_{i}}$, $\forall i \in \mathcal A, j\in \mathcal J_i$, (\ref{max1_3_2}b) can be replaced by
\begin{equation}\label{AlEq1}
\text{e}^{u_{j}-v_{i}+b_{j}}+\sum_{k\in\mathcal A \setminus \{i\}} \text{e}^{u_{j}+v_k-v_{i}+a_{kj}}\leq 1, \quad\forall i\in\mathcal A, j \in \mathcal J_i,
\end{equation}
where $a_{kj}=\ln\frac{  g_{kj}}{g_{ij}}$, and $b_{j}=\ln\frac{\sigma^2}{g_{ij}}$, $\forall j \in \mathcal J_i$.
Denote $w_{j}=u_{j}-v_{i}+b_{j}$, and $s_{ij}=u_{j}+v_i-v_{k}+a_{ij}, \forall i, k \in \mathcal A, i\neq k, j \in \mathcal J_k$.
Then, problem (\ref{max1_3_2}) is equivalent to
\begin{subequations}\label{max1_4}
\begin{align}
\!\!\!\!\!\!\!\!\!\!\!\!\!\!\!
\mathop{\max}_{\pmb{u}, \pmb{v}, \pmb w, \pmb s} \quad
&\sum_{i\in\mathcal A}\sum_{j \in {\cal{J}}_i} \omega_j \ln\left(\ln(1+\text{e}^{u_{j}}) \right)
\\
\textrm{s.t.} \quad\;\;
&\text{e}^{w_{j}}+\sum_{k\in\mathcal A \setminus \{i\}} \text{e}^{s_{kj}}\leq 1,\quad \forall i \in \mathcal A, j \in \mathcal J_i\\
&w_{j}=u_{j}-v_{i}+b_{j}, \quad\forall i \in \mathcal A, j \in \mathcal J_i\\
&s_{ij}\!=\!u_{j}\!+\!v_i\!-\!v_{k}\!+\!a_{ij}, \!\!\quad\forall i, k \in \mathcal A, i\neq k, j \in \mathcal J_k\!\!\!\\
&{v_{i}}\leq \ln (P_i), \quad\forall i \in \mathcal A,
\end{align}
\end{subequations}
where $\pmb u=\{u_j\}_{j\in\mathcal J}$,
$\pmb v=\{v_i\}_{i\in\mathcal A}$, $\pmb w=\{w_j\}_{j\in\mathcal J}$, and $\pmb s=\{s_{ij}\}_{i,k \in\mathcal A, i\neq k, j\in\mathcal J_k}$.

Since
\begin{equation*}
\frac{\partial^2 \ln(\ln(1+\text{e}^{u_{j}}))}{\partial u_{j}^2}=\frac{\text{e}^{u_{j}}(\ln(1+\text{e}^{u_{j}})-\text{e}^{u_{j}})}
{(1+\text{e}^{u_{j}})^2\ln^2(1+\text{e}^{u_{j}})},
\end{equation*}
and $\ln (1+\text{e}^{u_{j}})-\text{e}^{u_{j}}<0$,
the objective function (\ref{max1_4}a) is a concave function.
Further considering the fact that constraints of problem (\ref{max1_4}) are all convex,
problem (\ref{max1_4}) is a convex problem, which can be effectively solved by the well-established methods \cite{boyd2004convex}.

Instead of using the interior-point method, we here adopt the dual method with low complexity to obtain the optimal solution to problem (\ref{max1_4}).
The Lagrangian function of problem (\ref{max1_4}) is
\begin{eqnarray}\label{Aeq6}
\mathcal L_2(\pmb u, &&\!\!\!\!\!\!\!\!\!\!\pmb v, \pmb w,\pmb s,\pmb \alpha,\pmb \beta, \pmb \lambda, \pmb \zeta)
=
\sum_{i\in \mathcal A}\sum_{j \in {\cal{J}}_i} \omega_j \ln\left(\ln(1+\text{e}^{u_{j}}) \right)
\nonumber \\
&&\!\!\!\!\!\!\!\!\!\!\!\!\!\!\!\!
-\sum_{i\in \mathcal A}\sum_{j \in {\cal{J}}_i} \alpha_{j}\left(\text{e}^{w_{j}}+\sum_{k\in\mathcal A \setminus \{i\}} \text{e}^{s_{kj}}-1\right)
\nonumber \\
&&\!\!\!\!\!\!\!\!\!\!\!\!\!\!\!\!
-\!\sum_{i\in \mathcal A}\sum_{j \in {\cal{J}}_i} \beta_j (w_{j}\!-\!u_{j}\!+\!v_{i}\!-\!b_{j})\!-\!\sum_{i \in \mathcal A} \zeta_i({v_{i}}\!- \!\ln( P_i))
\nonumber \\
&&\!\!\!\!\!\!\!\!\!\!\!\!\!\!\!\!
\!-\!\sum_{i, k \in \mathcal A, i\neq k}
\sum_{j \in {\cal{J}}_k}\lambda_{ij} (s_{ij}\!-\!u_{j}\!-\!v_i\!+\!v_{k}\!-\!a_{ij})
,\nonumber
\end{eqnarray}
where $\pmb \alpha=\{\alpha_{i}\}_{j\in\mathcal J}$,
$\pmb \beta=\{\beta_{j}\}_{j\in\mathcal J}$,
$\pmb \lambda=\{\lambda_{ij}\}_{i,k \in \mathcal A, i\neq k, j\in\mathcal J_k}$,
and $\pmb \zeta=\{\zeta_{i}\}_{i\in\mathcal A}$.
$\pmb \alpha \geq \pmb 0$, $\pmb \beta$, $\pmb \lambda$ and $\pmb \zeta\geq \pmb 0$ are Lagrange multipliers associated with the corresponding constraints of problem (\ref{max1_4}).

\begin{algorithm}[h]
\caption{Iterative user association, load distribution and power control (IULP) algorithm}
\label{alg:Framwork1}
\begin{algorithmic}[1]
\State Initialize any feasible solution ($\pmb x^{(0)}$, $\pmb d^{(0)}$, $\pmb p^{(0)}$) of problem (\ref{max1_2}), the tolerance $\xi$, the iteration number $t=1$, and the maximal iteration number $T_{\max}$.
\State Compute objective value $V_{\text{obj}}^{(0)}\!=\!\bar V(\pmb x^{(0)}, \pmb d^{(0)}, \pmb p^{(0)})$, where $\bar V(\pmb x, \pmb d, \pmb p)\!=\!\sum_{i \in \mathcal I}\!\sum_{j \in {\cal{J}}}\!\omega_j x_{ij}\log_2\left(\frac{ KB\omega_j d_i \log_2(1+\eta_{ij})}{\sum_{l\in \mathcal J}\omega_l x_{il}}\right)$.
\State Obtain the optimal $\pmb x^{(t)}$ of problem (\ref{max1_2}) with fixed  $(\pmb d^{(t-1)}, \pmb p^{(t-1)})$ by solving (\ref{max1_2_21}).
\State Obtain the optimal  $(\pmb d^{(t)}, \pmb p^{(t)})$ of problem (\ref{max1_2}) with fixed $\pmb x^{(t)}$ by solving (\ref{max1_3_2}).
\State
Compute objective value $V_{\text{obj}}^{(t)}=\bar V(\pmb x^{(t)}, \pmb d^{(t)}, \pmb p^{(t)})$.
If $\left|V_{\text{obj}}^{(t)}-V_{\text{obj}}^{(t-1)}\right|\Big/V_{\text{obj}}^{(t-1)}<\xi$ or $t>T_{\max}$, output $\pmb x^* =\pmb x^{(t)}, \pmb d^*={\pmb d^{(t)}}, \pmb p^*={\pmb p^{(t)}}$ and terminate.
Otherwise, set $t=t+1$ and go to step 3.
\end{algorithmic}
\end{algorithm}

By using the dual method, the optimal solution to problem (\ref{max1_4}) is obtained by iteratively optimizing primal variables ($\pmb u, \pmb v, \pmb w, \pmb s$) with fixed dual variables ($\pmb \alpha, \pmb\beta, \pmb \lambda, \pmb\zeta$), and updating dual variables ($\pmb \alpha, \pmb\beta, \pmb \lambda, \pmb\zeta$) with fixed primal variables ($\pmb u, \pmb v, \pmb w, \pmb s$).
The details are given in Appendix C.

\subsection{Iterative User Association, Load Distribution and Power Control Algorithm}
We present the iterative algorithm in Algorithm 1, which is referred as iterative user association, load distribution and power control (IULP) algorithm.

\begin{theorem}
Assuming $T_{\max}\rightarrow \infty$, the sequence ($\pmb x$, $\pmb d$, $\pmb p$) generated by IULP algorithm converges.
\end{theorem}
\itshape \textbf{Proof:}  \upshape Please refer to Appendix D. \hfill $\Box$

\section{Optimization with Inner-Cell Unequal Power Allocation}
In Section $\text{\uppercase\expandafter{\romannumeral2}}$, we assume that each BS transmits with the same power for the associated users.
However, the equal power control strategy could be less efficient especially when the channel gains for different users associated with the same BS are quite different.
According to the well-known textbook \cite[Section 5.3.3]{tse2005fundamentals} as well as the literatures, e.g., \cite{Wong1999Multiuser,983324,1177182,Andrews2005Adaptive,1258257}, the system capacity can be enhanced with unequal power allocation.
In the following, we consider the case that each BS transmits with different power for different users.

Denote ($\pmb x^*, \pmb d^*, \pmb p^*$) as the solution to IULP algorithm.
Let $\mathcal J_i=\{ j\in \mathcal J|x_{ij}^*=1\}$ and $\mathcal A=\{i\in\mathcal I |d_i^*=1\}$ be the set of users associated with BS $i$ and the BS set operating at full load, respectively.
The load $d_i^*$ of BS $i$ is defined to be the proportion of time-frequency resource blocks consumed in BS $i$ due to serving all users in $\mathcal J_i$.
According to Theorem~2, $ d_i^*\in\{0, 1\}$, $\forall i \in \mathcal I$.
When $d_i^*=\sum_{j\in\mathcal J_i}y_{ij} = 0$, we can obtain $y_{ij} = 0$, $\forall j \in \mathcal J_i$.
Thus, we only need to consider the resource allocation and power control of BSs in $\mathcal A$.

We use $p_{ij}$ to denote the allocated power to user $j\in\mathcal J_i$ associated with BS $i$ on each resource block.
Assume that the average transmit power of BS $i \in \mathcal A$ on each resource block is fixed as:
\begin{equation}\label{FurEq1}
\sum_{i\in\mathcal J_i}y_{ij} p_{ij}=p_i^*, \quad\forall i \in \mathcal A.
\end{equation}
The SINR expression of user $j\in\mathcal J_i$ is
\begin{equation}\label{FurEq2}
\eta_{ij}=\frac{p_{ij} g_{ij}} {\sum_{k \in {\cal{I}}\setminus \{i\}}d_k^* p_{k}^* g_{kj} +\sigma ^2}, \quad\forall i \in \mathcal A, j \in \mathcal J_i.
\end{equation}
Since $ d_k^*\in\{0, 1\}$ and $p_{k}^*=0$ for $d_{k}^*=0$, we can obtain $\sum_{k \in {\cal{I}}\setminus \{i\}}d_k^* p_{k}^* g_{kj}={\sum_{k \in {\cal{A}}\setminus \{i\}} p_{k}^* g_{kj}}$.

With fixed user association, load distribution and average transmit power of each BS,
it is ready to formulate the utility maximization problem as
\begin{subequations}\label{Furmax1}
\begin{align}
\mathop{\max}_{\bar{\pmb{y}}, \bar {\pmb{p}}}
&\sum_{i\in \mathcal A}\sum_{j \in {\cal{J}}_i}\omega_j \log_2\left( KBy_{ij}\log_2\left(1+\frac{p_{ij}g_{ij}}{I_{ij}}\right)\right)
\\
\textrm{s.t.}\quad\!\!\!\!
&\sum_{j\in\mathcal J_i}y_{ij} = 1,\quad \forall i \in \mathcal A\\
&\sum_{i\in\mathcal J_i}y_{ij} p_{ij}=p_i^*, \quad\forall i \in \mathcal A\\
& y_{ij}\geq 0, p_{ij} \geq 0, \quad\forall i  \in {\cal{A}}, j \in \mathcal J_i,
\end{align}
\end{subequations}
where $\bar{\pmb y}=\{y_{ij}\}_{i \in \mathcal A, j \in \mathcal J_i}$, $\bar {\pmb p}=\{p_{ij}\}_{i \in \mathcal A, j \in \mathcal J_i}$ and $I_{ij}={\sum_{k \in {\cal{A}}\setminus \{i\}} p_{k}^* g_{kj} +\sigma ^2}$, $\forall i \in \mathcal I$, $j \in \mathcal J_i$.

According to the assumption that the average transmit power of each BS is fixed, the resource allocation and power control of each BS is independent with other BSs.
Due to the nonlinear constraints in (\ref{Furmax1}c), we introduce new power variables $q_{ij}=y_{ij}p_{ij}$, $\forall i \in \mathcal A, j \in \mathcal J_i$.
Then, the utility maximization problem of BS $i \in \mathcal A$ can be equivalent to
\begin{subequations}\label{Furmax2}
\begin{align}
\mathop{\max}_{ {\pmb{y}}_i, {\pmb{q}_i}}
&\sum_{j \in {\cal{J}}_i}\omega_j\ln\left( y_{ij}\ln\left(1+\frac{q_{ij}g_{ij}}{I_{ij}y_{ij}}\right)\right)
\\
\textrm{s.t.}\quad\!\!\!\!
&\sum_{j\in\mathcal J_i}y_{ij} = 1\\
&\sum_{i\in\mathcal J_i}q_{ij}=p_i^*\\
&y_{ij}\geq 0, q_{ij}\geq 0, \quad\forall j \in \mathcal J_i,
\end{align}
\end{subequations}
where $  {\pmb y}_i=\{y_{ij}\}_{j \in \mathcal J_i}$, and $ {\pmb q}_i=\{q_{ij}\}_{j \in \mathcal J_i}$.

Obviously, $g(q_{ij})\triangleq\ln\left(1+\frac{q_{ij}g_{ij}}{I_{ij}}\right)$ is concave with respect to (w.r.t.) $q_{ij}$.
According to the property of perspective function \cite[Section~3.2.6]{boyd2004convex},
$\bar g(y_{ij}, q_{ij})\triangleq y_{ij}g\left(\frac{q_{ij}}{y_{ij}}\right)
=y_{ij}\ln\left(1+\frac{q_{ij}g_{ij}}{I_{ij}y_{ij}}\right)$ is concave w.r.t. ($y_{ij}, q_{ij}$).
Since (\ref{Furmax2}a) is a nonnegative weighted sum of concave functions,
we can find that  (\ref{Furmax2}a) is concave w.r.t. (${\pmb{y}}_i, {\pmb{q}_i}$).
As a result, problem (\ref{Furmax2}) is a convex problem, of which the globally optimal solution can be obtained by using the dual method.

The Lagrange function in problem (\ref{Furmax2}) can be written by
\begin{eqnarray}\label{FurAeq6}
\mathcal L_3(\pmb y_i, &&\!\!\!\!\!\!\!\!\!\!\pmb q_i, \psi_i, \phi_i)
=
\sum_{j \in {\cal{J}}_i}\omega_j\ln\left( y_{ij}\ln\left(1+\frac{q_{ij}g_{ij}}{I_{ij}y_{ij}}\right)\right)
\nonumber \\
&&\!\!\!\!\!\!\!\!\!\!\!\!\!\!\!\!\!\!
-\psi_i\left(\sum_{j\in\mathcal J_i}y_{ij}-1\right)
- \phi_i \left(\sum_{i\in\mathcal J_i}q_{ij}-p_i^*\right),\nonumber
\end{eqnarray}
where $\psi_i$ and $\phi_i$ are the Lagrange multipliers associated with  (\ref{Furmax2}b) and
(\ref{Furmax2}c), respectively.
To solve convex problem (\ref{Furmax2}) with the dual method,
the details of optimizing the primal variables with fixed dual variables and updating the dual variables under given optimized primal variables are provided in Appendix E.

\section{Complexity Analysis}

\subsection{User Association with Unequal User Priorities}

To obtain the optimal user association via DGP algorithm (\ref{sol1})-(\ref{sol3}), the complexity is $\mathcal O(L_{\text{DG}}IJ)$, where $L_{\text{DG}}$ is the average number of iterations by the gradient projection method.
Regarding approximate belief propagation (ABP) algorithm to obtain a near optimal solution in \cite{chenuser}, the main computational complexity lies in the computation of message passed from the BS to users, which involves $\mathcal O(J)$ combinations.
Since in each iteration every BS passes the message, the total complexity of ABP algorithm is
$\mathcal O(L_{\text{AB}}IJ)$, where $L_{\text{AB}}$ denotes the total iteration number.
According to \cite{chenuser}, the optimal user association can be obtained by using exact belief propagation (EBP) algorithm.
From \cite{chenuser}, the complexity of EBP is $\mathcal O(L_{\text{EB}}IJ2^J)$, where $L_{\text{EB}}$ denotes the total iteration number of EBP algorithm.
Note that the user association algorithms ABP and EBP in \cite{chenuser} as well as the DGP are all distributed.

\subsection{Joint User Association, Load Distribution and Power Control}

For the proposed IULP algorithm, the major complexity lies in solving two subproblems: user association problem, load distribution and power control problem.
The user association problem is solved by DGP algorithm.
For the load distribution and power control problem, the algorithm presented in Section \ref{loadpowerFixedAssociation} is denoted by LDPC algorithm.
In LDPC algorithm, the complexity of computing $\pmb u(t+1)$ from (\ref{kkt1_1}) is $\mathcal O(IJ\log_2(1/\epsilon_1))$, where $\mathcal O(\log_2(1/\epsilon_1))$ is the complexity of using bisection method to compute the inverse function $f^{-1}(\cdot)$ for accuracy $\epsilon_1$.
Then, the complexity of LDPC algorithm is $\mathcal O(L_{\text{LD}}I J\log_2(1/\epsilon_1))$, where $L_{\text{LD}}$ is the average number of iterations by using LDPC algorithm.
Thus, the total complexity of IULP algorithm is
$\mathcal O(L_{\text{IU}}L_{\text{DG}}IJ+
L_{\text{IU}}L_{\text{LD}}IJ\log_2(1/\epsilon_1))$, where $L_{\text{IU}}$ is the average number of iterations by using IULP algorithm.
According to \cite[Page 390]{bertsekas2009convex}, a sharp estimate of both $L_{\text{DG}}$ and $L_{\text{LD}}$ can be expressed as $\mathcal O(1/\sqrt{\epsilon_0})$, where $\epsilon_0$ is the accuracy of the dual method.
In this section, we set $\epsilon_0$ as the accuracy of all dual methods.
Thus, the total complexity of IULP algorithm can be further simplified as
$\mathcal O(L_{\text{IU}} IJ\log_2(1/\epsilon_1)/\sqrt{\epsilon_0})$.
As for iterative BS association and power control (IBAPC) algorithm in \cite{shen2014distributed}, we know from \cite{shen2014distributed} that this algorithm consists of solving two subproblems: BS association problem solved by using the dual coordinate descent method, and power control problem solved with Newton's method.
According to \cite{shen2014distributed}, each iteration has a complexity of $\mathcal O(L_{\text{DC}}IJ+L_{\text{NM}}I^2J)$, where $L_{\text{DC}}=\mathcal O(1/\sqrt{\epsilon_0})$ and $L_{\text{NM}}$ are the numbers of iterations required by the dual coordinate descent method and Newton's method, respectively.
Denoting $L_{\text{IB}}$ as the number of iterations required in IBAPC algorithm, IBAPC algorithm has a total complexity of $\mathcal O(L_{\text{IB}}L_{\text{DC}}IJ+L_{\text{IB}}L_{\text{NM}}I^2J)$.
Note that the power control step in both IBAPC and the proposed IULP is centralized according to \cite[Appendix~B]{shen2014distributed} and Appendix C.

\subsection{Inner-Cell Unequal Power Allocation}

To further exploit the multiuser gain, users associated with the same BS are allocated with different fractional resource blocks and transmit power according to the proposed inner-cell unequal power allocation (ICUPA) algorithm in Section IV.
For ICUPA algorithm, the major complexity lies in the computation of $y_{ij}(t)$ in each iteration.
The complexity of computing $y_{ij}(t)$ from (\ref{FurKKT4_3_1}) by using the bisection method is $\mathcal O(\log_2(1/\epsilon_2) )$ for accuracy $\epsilon_2$.
Thus, the total complexity of ICUPA algorithm is
$\mathcal O(L_{\text{IC}}IJ\log_2(1/\epsilon_2))$, where $L_{\text{IC}}=\mathcal O(1/\sqrt{\epsilon_0})$ denotes the total iteration number in the outer layer of ICUPA algorithm.
Table $\text{\uppercase\expandafter{\romannumeral1}}$ summarizes the complexity analysis.

The key parameters in Table I are $\epsilon_0$, $L_{\text{AB}}$, $L_{\text{EB}}$, $L_{\text{IU}}$, $L_{\text{IB}}$, $L_{\text{NM}}$, $\epsilon_1$, and $\epsilon_2$.
According to Section VI in \cite{chenuser}, the belief propagation algorithm converges within a few, e.g., five, iterations from simulations, i.e., typical values for the iteration numbers of ABP algorithm and EBP algorithm are $L_{\text{AB}}=5$ and $L_{\text{EB}}=5$, respectively.
According to Fig.~8 in the following Section VI, typical values of iteration numbers of the proposed IULP algorithm and the IBAPC algorithm in \cite{shen2014distributed} are $L_{\text{IU}}=5$ and $L_{\text{IB}}=5$, respectively.
From the theory of convex optimization \cite[Page~495]{boyd2004convex}, a typical iteration number to achieve very high accuracy $(10^{-5})$ for Newton's method is 18, i.e., $L_{\text{NM}}=18$.
Since $\epsilon_0$ represents the accuracy of the dual method and $\epsilon_1$ and $\epsilon_2$ represent the accuracy of the bisection method, we can typically set $\epsilon_0=\epsilon_1=\epsilon_2=10^{-5}$.

With the above elaborated typical parameters $1/\sqrt{\epsilon_0}=317$, $L_{\text{AB}}=L_{\text{EB}}=L_{\text{IU}}=L_{\text{IB}}=5$, $L_{\text{NM}}=18$ and $\log_2(1/\epsilon_1)=\log_2(1/\epsilon_2)=17$,
Table I presents the computational complexity for various algorithms using these typical values.
From Table I, it is observed that the proposed DGP has almost the same complexity as ABP, and it has much smaller complexity compared to EBP.
It is also found that the proposed IULP and ICUPA algorithms have almost the same complexity as IBAPC algorithm.
\begin{table*}[!htb]
\centering
\caption{Computational Complexity} \label{tab:complexity}
\begin{tabular}{ccc}
  \hline
  \hline
  Algorithm & Complexity & Complexity with Typical Values \\ \hline
  proposed DGP  & $\mathcal O(IJ/\sqrt{\epsilon_0})$ &  $\mathcal O(317IJ)$\\
  ABP \cite{chenuser}& $\mathcal O(L_{\text{AB}}IJ)$& $ \mathcal O(5IJ)$\\
  EBP \cite{chenuser}& $\mathcal O(L_{\text{EB}}IJ2^J)$& $\mathcal O(5IJ2^J)$\\
 proposed IULP &  $\mathcal O(L_{\text{IU}}IJ \log_2(1/\epsilon_1)/\sqrt{\epsilon_0})$ &   $\mathcal O(26945 IJ )$  \\
IBAPC \cite{shen2014distributed}&$\mathcal O(L_{\text{IB}}IJ/\sqrt{\epsilon_0}+L_{\text{IB}}L_{\text{NM}}I^2J)$& $\mathcal O(1585IJ+ 90I^2J)$\\
 proposed ICUPA   & $\mathcal O(IJ \log_2(1/\epsilon_2)/\sqrt{\epsilon_0})$ & $\mathcal O(5389 IJ)$ \\
  \hline
  \hline
\end{tabular}
\end{table*}

\section{Simulation Results}

In this section,
we present simulation results for the proposed DGP, IULP and ICUPA algorithms.
We consider a three-tier HetNet with one macro BS (MBS), two pico BSs (PBSs) and two femto BSs (FBSs), as shown in Fig. 2.
The transmit power of the three-tier HetNet is $\{46, 38, 30\}$ dBm.
We assume that there are a total number of 50 users uniformly distributed in the HetNet.
A number of 20 users randomly chosen from the 50 users are set with higher priority, i.e., $\omega_j=2$, and the remaining 30 users are set with lower priority, i.e., $\omega_j=1$.
The total number of time-frequency resource blocks is $55$ for each BS, and the bandwidth of each time-frequency resource is $B=180$ KHz.
The noise power is $\sigma^2=-104$ dBm and the parameter $T$ in  (\ref{max3_2}) in Appendix C is set as $10^{-3}$.
In modeling the propagation environment, we respectively use the large-scale path loss $L(d)=34+40\log(d)$ and $L(d)=37+30\log(d)$ for MBS/PBSs and FBSs \cite{ye2013user}, where $d$ is measured in meter.
Besides, the standard deviation of shadow fading is set as $8$ dB.
Note that the simulation plots have been smoothed by averaging over 1000 realizations.

\begin{figure}
\centering
\includegraphics[width=3.2in]{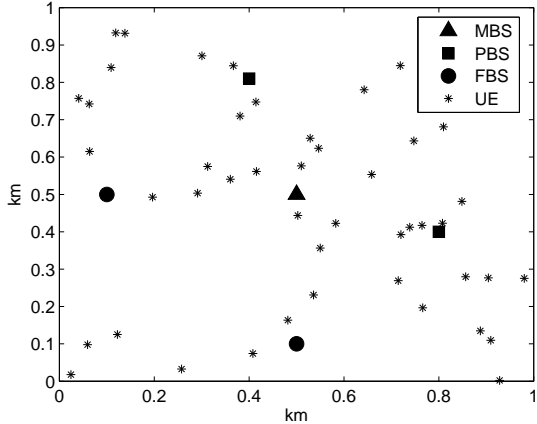}
\vspace{-1em}
\caption{Network configuration and user distribution of a three-tier HetNet.}
\end{figure}

We compare our proposed algorithms with a number of existing algorithms as follows:
\begin{itemize}
  \item DGP-MP: the proposed DGP algorithm under the maximal transmit power;
  \item MSINR-MP: the max-SINR association under the maximal transmit power;
  \item ABP-MP: the existing ABP algorithm \cite{chenuser} under the maximal transmit power;
  \item IULP: the proposed iterative user association, load distribution and power control algorithm;
  \item IBAPC: the iterative BS association and power control algorithm in \cite{shen2014distributed};
  \item DDO: the direct dual optimization algorithm
  \cite[Section IV-B]{shen2014distributed} with many initial points to better approach the global optimum;
  \item MSINR-MP+ICUPA: run the proposed inner-cell unequal power allocation (ICUPA) algorithm after MSINR-MP;
  \item IULP+ICUPA: run the proposed ICUPA algorithm after IULP.
\end{itemize}

\subsection{User Association under Fixed Load Distribution and Power Control}

\begin{table}[!htb]
\centering
\caption{Utility Values for Various User Association Algorithms}
\begin{tabular}{cccc}
  \hline
    & MSINR-MP&DGP-MP&ABP-MP  \\ \hline
 Utility  &26.02  &40.43 &32.06 \\
  \hline
\end{tabular}
\end{table}

\begin{figure}
\centering
\includegraphics[width=3.2in]{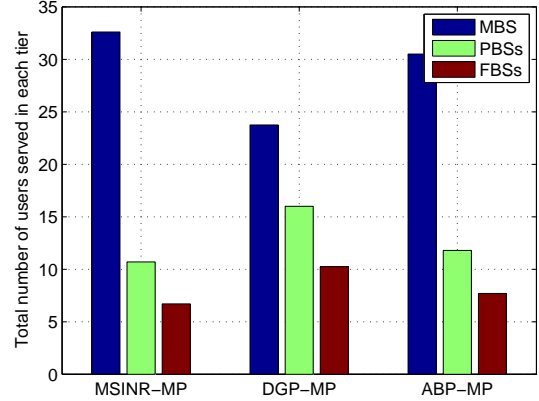}
\vspace{-1em}
\caption{Comparison of total number of users per tier for different user association algorithms.}
\end{figure}

Table II shows that the numerical utility\footnote{The numerical value of the utility is computed as the sum of log of user rates where rates are measured in Mbits/s.} achieved by MSINR-MP, DGP-MP and ABP-MP.
According to Table I and Table II, we can observe that the complexity of ABP-MP and DGP-MP is almost the same, but DGP-MP has presented better sum utility performance than ABP-MP.
Fig.~3 compares the total number of users per tier for different user association algorithms.
Obviously, MSINR-MP results in the largest total number of users served by MBS and the smallest number of users served by FBSs when compared with the other two algorithms.
In DGP-MP, many users are shifted to the less congested FBSs and PBSs, which suggests that our objective function can realize rate fairness.

Fig. 4 shows the cumulative distribution functions (CDFs) of user rates in the HetNet with different user association algorithms.
The CDF of user rate is evaluated as $F(r)=\Pr(R\leq r)$, where $\Pr(\cdot)$ is the probability function and $R$ is the user rate.
Denote the inverse function of $F(r)$ by $F^{-1}(p)$, where $p$ is the probability variable.
The gains of rate with $p$ for various user association algorithms vs. MSINR-MP are presented in Fig. 5.
The rate gain function $g(p)$ for a given probability vs. MSINR-MP is defined as
\begin{equation}
g(p)=\frac{F^{-1}_{\text{X}} (p)}
{F^{-1}_{\text{MSINR-MP}} (p)},
\end{equation}
where $F^{-1}_{\text{MSINR-MP}} (p)$ is the inverse CDF of MSINR-MP, and $F^{-1}_{\text{X}}(p)$ is the inverse CDF of (X=)DGP-MP or (X=)ABP-MP.
From Fig. 5, it can be observed that the rate gain for the proposed DGP-MP improves significantly at low rate vs. MSINR-MP, while the rate gain for ABP-MP improves slightly at low rate vs. MSINR-MP.
This is because the ABP-MP in \cite{chenuser} reduces the computation with approximation, while the DGP-MP can obtain the optimal solution without approximation.
However, for moderate-rate users, the ABP-MP slightly outperforms our proposed DGP-MP.
This is due to that the DGP-MP can guarantee fairness for users with low rate at the sacrifice of reducing rates of moderate-rate users.
According to Table II, the sum utility of the proposed DGP-MP is almost 26\% larger than ABP-MP, which demonstrates that the average performance of the DGP-MP is superior over the ABP-MP.

Note that the rate gain at low probability level in \cite{ye2013user} is over 3 but the rate gain at low probability level in our paper is less than that due to the following two reasons.
According to Fig.~6 in \cite{ye2013user} and Fig.~5 in our paper, the rate gain at low probability level is equivalent to the rate gain for users with low rates (equivalently, users with low channel gains to some degrees).
The first reason is that our paper considers unequal user priorities and users with high channel gains can be assigned with high priorities, which results in low rate gain for users with low channel gains.
The second reason is about the user distribution model. \cite{ye2013user} models the location processes across different tiers as independent with deployed density $\{\lambda_2, \lambda_3\} = \{5, 20\}$ per macrocell, which assumes that a large number of users are distributed in the third tier, i.e., in the range of femtocells.
With this user distribution model, the Max-SINR rule results in a large number of low-rate users associated with the macro BS, which should be associated with the pico/femto BSs.
Thus, the rate gain for users with low rates of the user association algorithm vs. the Max-SINR can be high.
Our paper assumes that there are a total number of 50 users uniformly distributed in the network, which leads to lower rate gain for users with low rates in our paper than that in \cite{ye2013user}.

\begin{figure}
\centering
\includegraphics[width=3.2in]{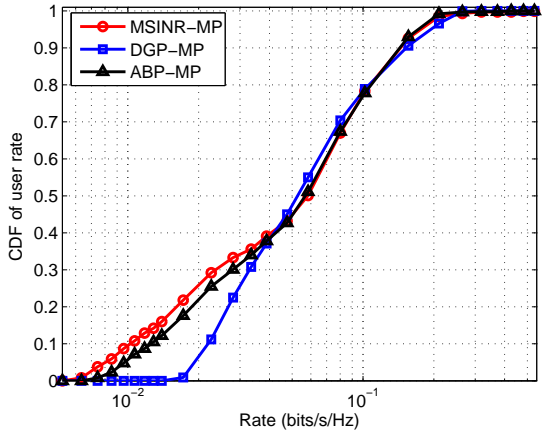}
\vspace{-1em}
\caption{CDFs of user rates for different user association algorithms.}
\end{figure}

\begin{figure}
\centering
\includegraphics[width=3.2in]{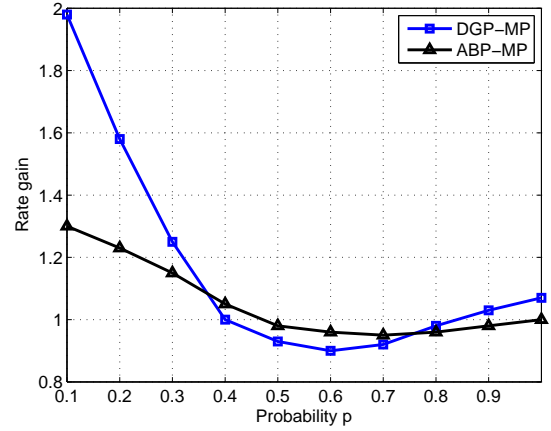}
\vspace{-1.0em}
\caption{Rate gains for different user association algorithms.}
\end{figure}

\subsection{Joint User Association, Load Distribution and Power Control}
\begin{table*}[!htb]
\centering
\caption{Utility Values for Various Joint User Association, Load Distribution and Power Control Algorithms}
\begin{tabular}{cccccc}
  \hline
    & IBAPC&DDO &IULP&MSINR-MP+ICUPA&IULP+ICUPA  \\ \hline
 Utility  &99.36 & 113.36& 105.05 & 30.63 &109.67  \\
  \hline
\end{tabular}
\end{table*}

\begin{figure}
\centering
\includegraphics[width=3.2in]{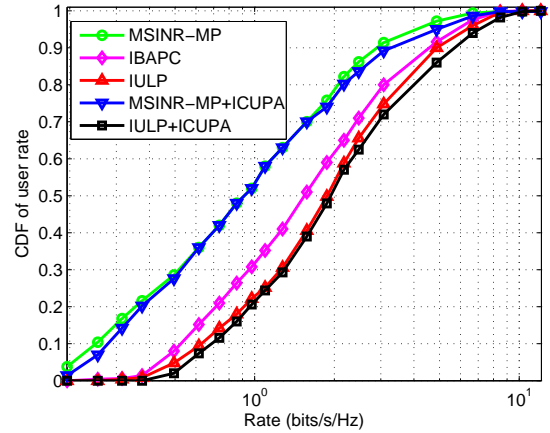}
\vspace{-1em}
\caption{ CDFs of user rates for different joint user association, load distribution and power control algorithms.}
\end{figure}

\begin{figure}
\centering
\includegraphics[width=3.2in]{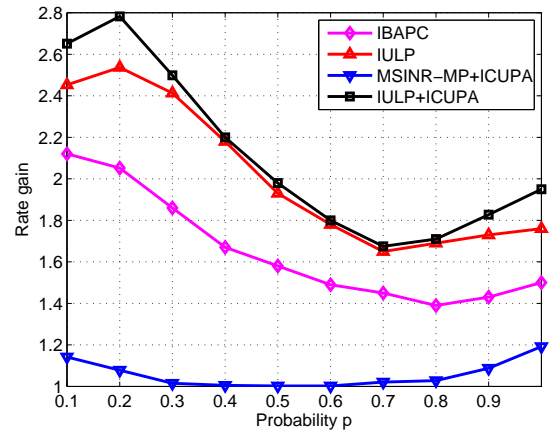}
\vspace{-1em}
\caption{Rate gains for different joint user association, load distribution and power control algorithms.}
\end{figure}

The numerical utility achieved by joint user association, load distribution and power control algorithms is compared in Table III.
Fig. 6 shows the CDFs of user rates for different joint user association, load distribution and power control algorithms.
From Fig. 6, Table II and Table III, we can observe significant difference between MSINR-MP and joint user association, load distribution and power control algorithms IULP and IBAPC.
Moreover, the gains of rate with $p$ for various joint user association, load distribution and power control algorithms vs. MSINR-MP are presented in Fig. 7.
The rate gain is quite large for IULP and IBAPC at low rate (e.g., 2.1-2.5x vs. MSINR-MP at the 10 \% rate point).
According to Fig.~6 and Fig.~7, we can find that IULP always outperforms IBAPC.
This is because the optimal solution to power control problem is obtained by solving an equivalent convex problem in IULP, and the suboptimal solution to power control problem is obtained by using Newton's method in IBAPC.
Using ICUPA, the performance of low-rate users and high-rate users can be further improved from Fig. 6 and Fig.~7.
Since channel conditions of users associated with the same BS are usually different, the multiuser gains can be realized in ICUPA by setting unequal power for different users associated with the same BS, especially for users with low channel gains and high channel gains.

\begin{figure}
\centering
\includegraphics[width=3.2in]{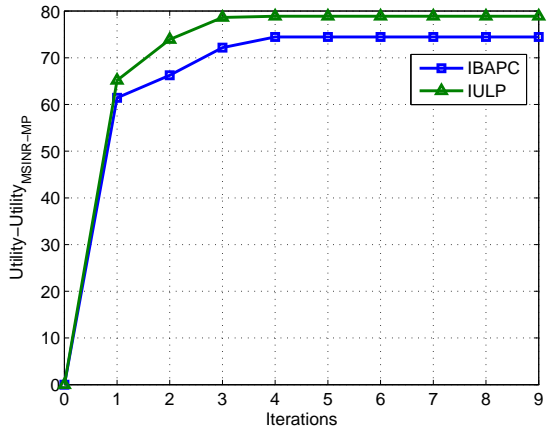}
\vspace{-1em}
\caption{Convergence behaviors for different joint user association, load distribution and power control algorithms.}
\end{figure}

\begin{figure}
\centering
\includegraphics[width=3.2in]{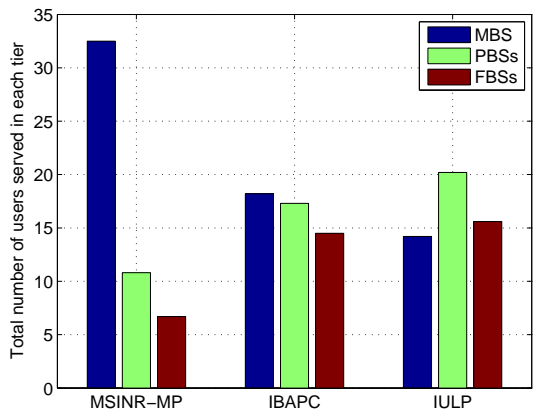}
\vspace{-1em}
\caption{Comparison of total number of users per tier for different joint user association, load distribution and power control algorithms.}
\end{figure}

\begin{figure}
\centering
\includegraphics[width=3.2in]{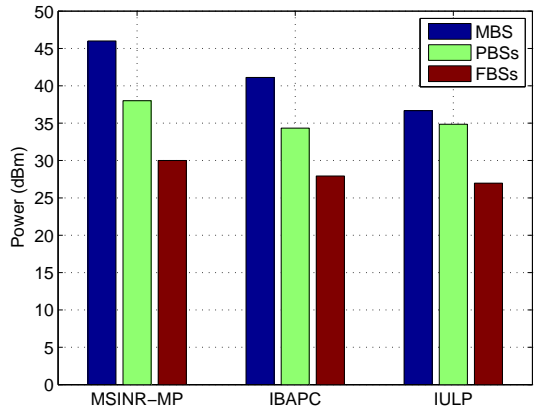}
\vspace{-1em}
\caption{Optimized average power levels in per tier for different joint user association, load distribution and power control algorithms.}
\end{figure}

Fig. 8 illustrates the convergence behaviors for IULP and IBAPC.
The y-axis of Fig. 8 means the utility difference which is the utility value of IBAPC (or IULP) subtracted by the utility value of MSINR-MP.
It can be seen that both IULP and IBAPC converge rapidly.
Obviously, the utility value by IULP outperforms IBAPC.
This is because the optimal solution to power control problem is obtained in IULP.

In Fig. 9, we show the total number of users per tier for different joint user association, load distribution and power control algorithms.
It is observed that the algorithms with better performance tend to have a larger number of users in PBSs and FBSs, which illustrates the benefit of offloading traffic from MBS to PBSs and FBSs.
In addition, Fig. 10 compares the power levels produced by various joint user association, load distribution and power control algorithms.
Since there are two PBSs and two FBSs, we plot the averaged transmit power of both PBSs and FBSs in Fig.~10.
Combing Fig.~9 and Fig. 10, we conclude that a combination of setting MBS with lower power and smaller number of associated users is the key to obtaining good system performance.

\section{Conclusion}

In this paper, we have investigated the logarithmic utility maximization problem for load-coupled HetNets with unequal user priorities.
By analyzing the utility function, we prove that it is optimal for each BS to operate at full load or zero load.
We propose an iterative algorithm, which consists
of solving two subproblems: the user association problem and the load distribution and power control problem.
The user association problem can be solved by using the dual gradient projection method.
Although the load distribution and power control problem with fixed user association is nonconvex, we transform it into an equivalent convex problem.
To further exploit the multiuser gain, we provide the logarithmic utility maximization problem for inner-cell unequal power allocation, and obtain the optimal solution by using the dual method.
Simulation results show that the proposed algorithm achieves better performance than conventional algorithms in terms of logarithmic utility.

\appendices

\section{}
Given ($\pmb x^*, \pmb d^*, \pmb p^*$) with $\pmb d^*=[d_1^*, \cdots, d_I^*]^T$, problem (\ref{max1_1}) becomes the following resource allocation problem
\begin{subequations}\label{apend2_1}
\begin{align}
\mathop{\max}_{ \pmb{y}}\quad
&\sum_{j \in {\cal{J}}}\omega_j \log_2\left(\sum_{i\in {\cal{I}}}y_{ij}KB\log_2(1+\eta_{ij}^*)\right)
\\
\textrm{s.t.}\quad\:
& \sum_{j\in {\cal{J}}}y_{ij}=d_i^*, \quad\forall i  \in {\cal{I}}\\
&0\leq y_{ij}\leq x_{ij}^*, \quad\forall i  \in {\cal{I}}, j \in \mathcal J,
\end{align}
\end{subequations}
where $\eta_{ij}^*=\frac{p_{i}^* g_{ij}} {\sum_{k \in {\cal{I}}\setminus \{i\}} d_k^* p_{k}^* g_{kj} +\sigma ^2}, \forall i  \in {\cal{I}}, j \in \mathcal J$.

If $d_i^*=0$, we have $y_{ij}^*=0$ from (\ref{apend2_1}b) and (\ref{apend2_1}c).
Thus, we only need to consider the case $d_i^*>0$.
Since $x_{ij}^*$ is a user association variable, $x_{ij}^*\in\{0,1\}$.
If $x_{ij}^*=0$, we can obtain $y_{ij}^*=0$ from (\ref{apend2_1}c).
Due to the fact that each user is associated with only one BS, the objective function of (\ref{apend2_1}) can be rewritten as
\begin{subequations}\label{apend2_2_2}
\begin{align}
&\sum_{i \in \mathcal I}\sum_{j \in \{l|x_{il}^*=1\}}\omega_j \log_2\left(y_{ij}KB\log_2(1+\eta_{ij}^*)\right)\nonumber\\
&\!\!\!\!\!\!\!\!\!
=
\sum_{i \in \mathcal I}\sum_{j \in \{l|x_{il}^*=1\}}\omega_j \log_2\left(y_{ij}\right)
\nonumber\\
&\!\!\!\!\!\!\!\!\!
\quad
+
\sum_{i \in \mathcal I}\sum_{j \in \{l|x_{il}^*=1\}}\omega_j \log_2\left(KB\log_2(1+\eta_{ij}^*)\right),
\end{align}
\end{subequations}
where the second term is constant.
The above manipulation decouples the problem in (\ref{apend2_1}) into a sequence of resource allocation optimizations per BS.
Then, we conduct the resource allocation on a typical BS $i$ with $d_i^*>0$ and the users associated with BS $i$.
The utility maximization problem for users associated with BS $i$ is
\begin{subequations}\label{apend2_2}
\begin{align}
\mathop{\max}_{ \pmb{y}_i}\quad
&\sum_{j \in \{l|x_{il}^*=1\}}\omega_j \log_2\left(y_{ij}\right)
\\
\textrm{s.t.}\quad\:
&\sum_{j\in\{l|x_{il}^*=1\}}y_{ij}=d_i^*\\
&0\leq y_{ij}\leq 1, \quad\forall j\in\{l|x_{il}^*=1\},
\end{align}
\end{subequations}
where $\pmb y_i=\{y_{ij}\}_{j\in\{l|x_{il}^*=1\}}$.
Obviously, resource allocation problem (\ref{apend2_2}) is convex and we can obtain the optimal solution by solving the Karush-Kuhn-Tucker (KKT) conditions.
Denoting by $\chi$ the Lagrange multiplier associated to (\ref{apend2_2}b), the Lagrange function of problem (\ref{apend2_2}) is
\begin{equation}\label{apend2_3}
\mathcal L_1(\pmb y_i, \chi)
\!=\!
\sum_{j \in \{l|x_{il}^*=1\}}\omega_j \log_2( y_{ij})
+\chi\left(\!\sum_{j\in\{l|x_{il}^*=1\}}y_{ij}-d_i^*\!\right).
\end{equation}
From \cite{boyd2004convex}, the optimal solution should satisfy the following KKT conditions of problem (\ref{apend2_2}):
\begin{equation}\label{apend2_4}
\frac{\mathcal L_1}{\partial y_{ij}}=
\frac{\omega_j}{(\ln 2)y_{ij}}-\chi=0, \quad\forall j\in\{l|x_{il}^*=1\},
\end{equation}
which yields
\begin{equation}\label{apend2_5}
y_{ij}=\frac{\omega_j}{(\ln 2)\chi}, \quad\forall j\in\{l|x_{il}^*=1\}.
\end{equation}
Substituting (\ref{apend2_5}) into (\ref{apend2_2}b), we can obtain
\begin{equation}\label{apend2_6}
\chi=\frac{\sum_{j\in\{l|x_{il}^*=1\}}{\omega_l}}{(\ln 2)d_i^*}.
\end{equation}
By inserting (\ref{apend2_6}) into (\ref{apend2_5}), we have
\begin{equation}\label{apend2_7}
y_{ij}=\frac{\omega_j d_i^*}{\sum_{j\in\{l|x_{il}^*=1\}}{\omega_l}}, \quad\forall j\in\{l|x_{il}^*=1\}.
\end{equation}
Hence, Theorem 1 is proved.

\section{}

Applying the optimal resource allocation in (\ref{eqresour}) to problem (\ref{max1_1}) yields
\begin{subequations}\label{Aeqmax1_2}
\begin{align}
\mathop{\max}_{\pmb{x}, \pmb d, \pmb{p}} \:\:
&
\sum_{i \in \mathcal I}\sum_{j \in {\cal{J}}}\omega_j x_{ij}\log_2\left(\frac{KB \omega_j d_i \log_2(1+\eta_{ij}) }{\sum_{l\in \mathcal J}\omega_l x_{il}}\right)
\\
\textrm{s.t.} \quad\!\!
&\eta_{ij}=\frac{p_{i} g_{ij}} {\sum_{k \in {\cal{I}}\setminus \{i\}} d_k p_{k} g_{kj} +\sigma ^2}, \quad\forall i  \in {\cal{I}}, j \in \mathcal J\\
&\sum_{i\in\mathcal I}x_{ij} = 1, \quad\forall j \in \mathcal J\\
&0 \leq d_i \leq 1, \quad\forall i \in \mathcal I\\
&0 \leq p_{i} \leq P_i, \quad\forall i \in \mathcal I\\
& x_{ij} \in\{ 0, 1 \}, \quad\forall i  \in {\cal{I}}, \forall j \in \mathcal J,
\end{align}
\end{subequations}
where $\pmb d =[d_1, \cdots, d_I]^T$.
The equivalence of (\ref{Aeqmax1_2}a) and (\ref{max1_1}a) follows from the fact that each user is associated with only one BS, i.e., there exists a BS $b(j)$ such that $x_{b(j)j}=1$, and $x_{kj}=0$ for all $k\in\mathcal I\setminus \{b(j)\}$.
From (\ref{max1_1}a), we can obtain
\begin{eqnarray}\label{AeqObj6a}
&&\!\!\!\!\!\!\!\!\!
\quad\omega_j\log_2\left(KB  \sum_{i\in\mathcal I}  y_{ij} \log_2(1+\eta_{ij}) \right)
\nonumber\\
&&\!\!\!\!\!\!\!\!\!
=\omega_j\log_2\left(\sum_{i\in\mathcal I}\frac{KB \omega_j x_{ij} d_i \log_2(1+\eta_{ij}) }{\sum_{l\in \mathcal J}\omega_l x_{il}}\right)
\nonumber\\
&&\!\!\!\!\!\!\!\!\!
=\omega_j\log_2\left( \frac{KB \omega_j x_{b(j) j}d_{b(j)} \log_2(1+\eta_{b(j) j}) }{\sum_{l\in \mathcal J}\omega_l x_{b(j) l}}\right)
\nonumber\\
&&\!\!\!\!\!\!\!\!\!
=\sum_{i\in\mathcal I}\omega_j x_{ij}\log_2\left( \frac{KB \omega_j x_{i j}d_{i} \log_2(1+\eta_{i j}) }{\sum_{l\in \mathcal J}\omega_l x_{i l}}\right),
\end{eqnarray}
where the first equality follows from (\ref{eqresour}) in Theorem 1, the second equality holds because $x_{b(j)j}=1$ and $x_{kj}\!=\!0$ for $k\in\mathcal I\setminus \{b(j)\}$, and the last equality follows from $x_{b(j)j}\!=\!1$, and $\omega_j x_{kj}\log_2\left(\! \frac{KB \omega_j x_{k j}d_{k} \log_2(1\!+\!\eta_{k j}) }{\sum_{l\in \mathcal J}\omega_l x_{k l}}\!\right)\!=\!0$ for $x_{kj}\!=\!0$, $k\!\in\!\mathcal I\setminus \{b(j)\}$.
Equation (\ref{AeqObj6a}) justifies the equality between (\ref{max1_1}a) and (\ref{Aeqmax1_2}a), i.e., the sum w.r.t. $i$ and $x_{ij}$ can be shifted outside the logarithm function.

Since constraints (\ref{Aeqmax1_2}c) and (\ref{Aeqmax1_2}f) are only determined by user association $\pmb x$, the load distribution and power control problem (\ref{Aeqmax1_2}) with fixed user association $\pmb x$ is
\begin{subequations}\label{Aeqmax1_3}
\begin{align}
\mathop{\max}_{\pmb d, \pmb{p}} \; \!\!
&
\sum_{i \in \mathcal I}\!\!\sum_{j \in{\cal{J}}\!\!}\omega_j x_{ij}\log_2\!\!\left(\!\!d_i \!\log_2\!\!\left(\!\!1\!+ \!\frac{p_{i} g_{ij}} {\sum_{k \in {\cal{I}}\!\setminus \! \{i\}}\!d_k p_{k} g_{kj} \!+\!\sigma ^2}\!\!\right)\!\!\! \right)\!\!+\!\!C
\\
\textrm{s.t.} \;\;
&0\leq d_i \leq 1, \quad\forall i \in \mathcal I \\
&0 \leq p_{i} \leq P_i, \quad \forall i \in \mathcal I,
\end{align}
\end{subequations}
where
\begin{equation}\label{Aeqmax1_3Constant}
C\triangleq \sum_{i \in \mathcal I}\sum_{j \in {\cal{J}}}\omega_j x_{ij}\log_2\left(\frac{KB\omega_j }{\sum_{l\in \mathcal J}\omega_l x_{il}}\right)
\end{equation}
is a constant.
If either of $p_i$ and $d_i$ is 0, then the other is assumed to be 0.
Based on this assumption, we have the following lemma.
\begin{lemma}
If there exists at least one user $j$ such that $x_{ij}=1$, the optimal $d^*_i$ of problem (\ref{Aeqmax1_3}) is $d^*_i=1$; otherwise $d^*_i =0$.
\end{lemma}
\itshape \textbf{Proof:}  \upshape
Introducing a set of new variables: $z_i=d_i p_i$, $\forall i\in \mathcal I$, we can reformulate (\ref{Aeqmax1_3}) as:
\begin{subequations}\label{Aeqmax1_3_2}
\begin{align}
\mathop{\max}_{\pmb d, \pmb{z}} \;\!
&
\sum_{i \in \mathcal I}\!\sum_{j \in{\cal{J}}\!}\omega_j x_{ij}\log_2\!\left(\!d_i \log_2\!\left(\!1\!+ \!\frac{z_{i} g_{ij}} {d_i(\sum_{k \in {\cal{I}}\!\setminus \! \{i\}}\!z_{k} g_{kj} \!+\!\sigma ^2)}\!\right)\!\right)
\\
\textrm{s.t.} \;\;
&0\leq d_i \leq 1, \quad\forall i \in \mathcal I \\
&0 \leq z_{i} \leq P_i d_i, \quad \forall i \in \mathcal I,
\end{align}
\end{subequations}
where $\pmb z=[z_1, \cdots, z_I]^T$.
To investigate the monotonic property of the objective function (\ref{Aeqmax1_3_2}a) w.r.t. $d_i$, we define function
\begin{equation}
\vartheta(x)=\ln\left(x\ln\left(1+\frac{a}{x}\right)\right), \quad \forall x\geq 0,
\end{equation}
where $a>0$ is a positive constant.
The first-order derivative of $\vartheta(x)$ is
\begin{equation}
\vartheta'(x)=\frac{\left(1+\frac{a}{x}\right)\ln\left(1+\frac{a}{x}\right) -\frac{a}{x}}
{x\left(1+\frac{a}{x}\right)\ln\left(1+\frac{a}{x}\right) }, \quad \forall  x\geq 0.
\end{equation}
Denoting $\bar \vartheta(x)=(1+x)\ln(1+x)-x$, $\forall x\geq 0$, we can obtain
\begin{equation}
\bar\vartheta'(x)=\ln({1+x})\geq0, \quad \forall x\geq 0.
\end{equation}
Hence, $\bar \vartheta(x)\geq \bar \vartheta(0)=0$, which shows that $\vartheta'(x)\geq0$, $\forall x\geq 0$.
As a result, $\vartheta(x)$ is an increasing function w.r.t. $x$, i.e., objective function (\ref{Aeqmax1_3_2}a) monotonically increases w.r.t. $d_i$.
With any given $\pmb z$, the optimal load of BS $i$ is always $d_i^*=1$ if there exists at least one user $j$ such that $x_{ij}=1$.
For the case that $x_{ij}=0$ for all $j\in\mathcal J$, i.e., there is no user associated with BS $i$, we must have $z_i^*=0$ and the optimal load of BS $i$ is $d_i^*=0$ to reduce inter-cell interference.
\hfill $\Box$

Note that the load distribution problem to minimize sum power was solved by using the contradictory method in \cite[Lemma~2]{ho2015power}.
According to Lemma 1, the optimal $d^*_i$ of problem (\ref{Aeqmax1_3}) with any given user association $\pmb x$ satisfies $d_i^*\in\{0,1\}$.
Since problem (\ref{Aeqmax1_3}) is equivalent to problem (\ref{max1_1}) with fixed user association, the optimal $d^*_i$ of problem (\ref{max1_1}) must satisfy $d_i^*\in\{0,1\}$.

\section{}
According to \cite{boyd2004convex}, the optimal solution should satisfy the following KKT conditions of problem (\ref{max1_4}):
\begin{subequations}\label{KKT1}
\begin{align}
&
\frac {\partial \mathcal L_2}{\partial u_j}\!\!=\!\!
\frac{\omega_j\text{e}^{u_{j}}}{(1\!+\!\text{e}^{u_{j}})\ln(1\!+\!\text{e}^{u_{j}})}\!+\! \beta_j\!+\!\!\!\sum_{i \in {\cal{A}}\setminus\{b(j)\}}\lambda_{ij}\!\!=\!\!0,\quad  \forall j \in \mathcal J\!\!\!\!
\\
&\frac {\partial \mathcal L_2}{\partial v_i}=
-\sum\limits_{j \in {\cal{J}}_i} \beta_j-
\zeta_i+\sum\limits_{k\in \mathcal A\setminus\{i\}}\sum\limits_{j \in {\cal{J}}_k}\lambda_{ij}
\nonumber\\&\qquad\qquad
-
\sum\limits_{k\in \mathcal A\setminus\{i\}}\sum\limits_{j \in {\cal{J}}_i}\lambda_{ij}
=0, \qquad\forall i  \in {\cal{I}}\\
&\frac {\partial \mathcal L_2}{\partial w_j}=
\alpha_{j}\text{e}^{w_{j}}-\beta_j=0, \quad\forall j \in \mathcal J\\
&\frac {\partial \mathcal L_2}{\partial s_{ij}}=
\alpha_{j}\text{e}^{s_{ij}}-\lambda_{ij}=0, \quad\forall i,k \in \mathcal A, i\neq k, j \in \mathcal J_k,\tag{\ref{KKT1}d}
\end{align}
\end{subequations}
where $b(j)$ is the BS that associates with user $j$, i.e., $j\in\mathcal J_{b(j)}$.

To solve convex problem (\ref{max1_4}), we use the dual method by iteratively updating the Lagrange multipliers and primary variables.
In the $(t+1)$-th iteration, we can calculate the primary variables with given the Lagrange multipliers $\pmb \alpha(t)$, $\pmb \beta(t)$, $\pmb \lambda(t)$ and $\pmb \zeta(t)$.
Based on (\ref{KKT1}a), we have
\begin{equation}\label{kkt1_11}
\beta_j(t) +\!\sum_{i \in {\cal{A}}\setminus\{b(j)\}}\!\lambda_{ij}(t) \!=\!-\! \frac{\omega_j\text{e}^{u_{j}(t+1)}}{(1\!+\!\text{e}^{u_{j}(t+1)})
\ln(1\!+\!\text{e}^{u_{j}(t+1)})}
\!<\!0.
\end{equation}

Define function $f(x)=\frac{\text{e}^{x}}{(1+\text{e}^x)\ln(1+\text{e}^x)}$, which can be proved decreasing.
Denote $f^{-1}(x)$ as the inverse function of $f(x)$, $\forall x >0$.
From (\ref{kkt1_11}), we have
\begin{equation}\label{kkt1_1}
u_j(t+1)=f^{-1}\left(-\frac{\beta_j(t)+\sum_{i \in {\cal{A}}\setminus\{b(j)\}}\lambda_{ij}(t)}{\omega_j}\right), \quad \forall j \in \mathcal J.
\end{equation}

Since Lagrange function $\mathcal L_2$ is a linear function w.r.t $v_i$, the optimal value of $v_i$ can not be directly obtained.
To solve this, we introduce a small positive constant $T>0$, and the objective function of problem (\ref{max1_4}a) can be modified as,
\begin{equation}\label{max3_2}
\mathop{\max}_{\pmb{u}, \pmb{v}, \pmb w, \pmb s} \quad\sum_{i\in\mathcal A}\sum_{j \in {\cal{J}}_i} \ln\left(\ln(1+\text{e}^{u_{j}}) \right)
+T \sum_{i\in \mathcal A}\ln(1+\ln(P_i )-v_i).
\end{equation}
Obviously, the modified problem (\ref{max1_4}) with new objective function (\ref{max3_2}) is also a convex problem.
Besides, if the value of $T$ is as small as possible, the optimal solution to the modified problem approximately equals to the optimal solution to original problem (\ref{max1_4}).
By solving the KKT conditions of the modified problem, we can obtain the value of $v_i(t\!+\!1)$ in closed form as
\begin{eqnarray}
v_i(t+1)=
\left.\left[
1+\ln(P_i )-
\frac{T}{E}\right]
\right|^{\ln (P_i)} , \quad i \in \mathcal A,
\label{kkt1_4}
\end{eqnarray}
where $E=\sum_{k\in \mathcal A\setminus\{i\}}\sum_{j \in {\cal{J}}_k}\lambda_{ij}(t)-\sum_{j \in{\cal{J}}_i}\beta_j(t)-\sum_{k\in \mathcal A\setminus\{i\}}\sum_{j \in {\cal{J}}_i}\lambda_{ij}(t)-
\zeta_i(t)$, and
$x|^y=\min(x,y)$.

In the following, we use the contradiction method to prove that $\alpha_{j}(t) \neq 0$.
If $\alpha_j(t)=0$, we have $\beta_j(t)=0$ and $\lambda_{ij}(t)=0$ according to (\ref{KKT1}c) and (\ref{KKT1}d).
Thus, $\beta_j(t)+\sum_{i \in {\cal{A}}\setminus\{b(j)\}}\lambda_{ij}(t)=0, \forall i,k \in \mathcal A, i\neq k, j \in \mathcal J_k,
$ which contradicts (\ref{kkt1_11}).
Due to the fact that $\alpha_{j}(t) \neq 0$, we can respectively obtain $w_{j}(t+1)$ and $s_{ij}(t+1)$ from (\ref{KKT1}b) and (\ref{KKT1}c), i.e.,
\begin{equation}\label{kkt1_2}
w_{j}(t+1)=\ln\left(\frac{\beta_j(t)}{\alpha_{j}(t)}\right),\quad \forall j \in \mathcal J,
\end{equation}
and
\begin{equation}\label{kkt1_3}
s_{ij}(t+1)=\ln\left(\frac{ \lambda_{ij}(t)}{\alpha_{j}(t)}\right), \quad\forall i,k \in \mathcal A, i\neq k, j \in \mathcal J_k.
\end{equation}

To update the dual variables with the primal variables obtained from
(\ref{kkt1_1}), (\ref{kkt1_4})-(\ref{kkt1_3}), we exploit the gradient based method \cite{bertsekas2009convex}.
The new values of the Lagrange multipliers are updated by
\begin{eqnarray}
&&\!\!\!\!\!\!\!\!\!\!\!\!\!
\alpha_{j}(t+1)=\Bigg [\alpha_{j}(t)+\delta(t) \text{e}^{w_{j}(t+1)}
\nonumber\\
&&\!\!\!\!\!\!\!\!\!
+\delta(t)\sum_{k\in\mathcal A \setminus \{i\}} \text{e}^{s_{kj}(t+1)}-\delta(t)\Bigg]^+ , \quad \forall i \in \mathcal A, j \in \mathcal J_i \label{eqnar1}
\\
&&\!\!\!\!\!\!\!\!\!\!\!\!\!
\beta_j(t+1)=\beta_j(t)+\delta(t)(w_{j}(t+1)-u_{j}(t+1)
\nonumber\\
&&\!\!\!\!\!
+v_{i}(t+1)-b_{j}), \quad\forall i \in \mathcal A, j \in \mathcal J_i \label{eqnar2}
\\
&&\!\!\!\!\!\!\!\!\!\!\!\!\!
\lambda_{ij}(t  +1) =\lambda_{ij}(t)+ \delta(t)(s_{ij}(t+1)- u_{j}(t +1)- v_i(t + 1)
\nonumber\\
&&\!\!\!\!\!
+ v_{k}(t + 1) - a_{ij}), \quad\forall i,k \in \mathcal A, i\neq k, j \in \mathcal J_k \label{eqnar3}
\\
&&\!\!\!\!\!\!\!\!\!\!\!\!\!
\zeta_i(t+1)=\left[\zeta_i(t)+\delta(t)({v_{i}}(t+1)- \ln(P_i))\right]^+
, \!\!\quad\forall i \in \mathcal A, \label{eqnar4}
\end{eqnarray}
where $\delta(t)$ is a dynamically chosen stepsize sequence,
and $[x]^+$ denotes $\max\{x,0\}$.

\section{}
The proof is established by showing that the sum utility value (\ref{max1_2}a) is nondecreasing when sequence ($\pmb x$, $\pmb d$, $\pmb p$) is updated.
According to the IULP algorithm, we have
\begin{eqnarray}\label{convergenceProofIULP}
V_{\text{obj}}^{(t-1)}
&&\!\!\!\!\!\!\!\!\!\!
=\bar V(\pmb x^{(t-1)}, \pmb d^{(t-1)}, \pmb p^{(t-1)})
\nonumber\\
&&\!\!\!\!\!\!\!\!\!\!
\overset{(\text a)}{\leq}
\bar V(\pmb x^{(t)}, \pmb d^{(t-1)}, \pmb p^{(t-1)})
\nonumber\\
&&\!\!\!\!\!\!\!\!\!\!
\overset{(\text b)}{\leq}
\bar V(\pmb x^{(t)}, \pmb d^{(t)}, \pmb p^{(t)})
=V_{\text{obj}}^{(t)},
\end{eqnarray}
where inequality (a) follows from that $\pmb x^{(t)}$ is the optimal user association of problem (\ref{max1_2}) with fixed load and power $(\pmb d^{(t-1)}, \pmb p^{(t-1)})$, and inequality (b) follows from that $(\pmb d^{(t)}, \pmb p^{(t)})$ is the optimal load and power of problem (\ref{max1_2}) with fixed user association $\pmb x^{(t)}$.
Thus, the sum utility is nondecreasing after the update of user association, load distribution and power control.

Furthermore, the sum utility value (\ref{max1_2}a) can be upper-bounded by
\begin{eqnarray}\label{convergenceProofIULP1_2}
V_{\text{obj}}^{(t)}\!=\!&&\!\!\!\!\!\!\!\!\!
\sum_{i \in \mathcal I}\sum_{j \in {\cal{J}}}\omega_j x_{ij}^{(t)}\log_2\!\left(\!\frac{KB \omega_j d_i^{(t)} }{\sum_{l\in \mathcal J}\omega_l x_{il}^{(t)}}\log_2\!\left(\!1+ \frac{p_{i}^{(t)} g_{ij}} {I_{ij}^{(t)}}\!\right)\!\right)
\nonumber\\
\overset{(\text c)}{\leq}\!&&\!\!\!\!\!\!\!\!\!
\sum_{i \in \mathcal I}\sum_{j \in {\cal{J}}}\omega_j x_{ij}^{(t)}\log_2\!\left(\!\frac{K B\omega_j   }{\sum_{l\in \mathcal J}\omega_l x_{il}^{(t)}}\log_2\!\left(\!1+ \frac{P_{i} g_{ij}} {\sigma ^2}\!\right)\!\right)
\nonumber\\
\overset{(\text d)}{\leq}\!&&\!\!\!\!\!\!\!\!\!
\sum_{i \in \mathcal I}\!\sum_{j \in {\cal{J}}}\!\omega_j \log_2\!\left(\!\frac{ KB\omega_j }{ \min_{l\in\mathcal J}\{\omega_l \}}\!\log_2\!\left(\!1\!+\! \frac{P_{i} g_{ij}} {\sigma ^2}\!\right)\!\right),\!\!\!\!\!
\end{eqnarray}
where $I_{ij}^{(t)}\triangleq\sum_{k \in {\cal{I}}\setminus \{i\}} d_k^{(t)} p_{k}^{(t)} g_{kj} +\sigma ^2$, inequality (c) is due to $0\leq d_i^{(t)}\leq 1$ and $0 \leq p_{i}^{(t)}\leq P_i$,
and inequality (d) holds because $0\leq x_{ij}^{(t)}\leq 1$ and $\sum_{l\in \mathcal J}\omega_l x_{il}^{(t)}\geq \min_{l\in\mathcal J}\{\omega_l\}$ when there exists an $l$ such that $x_{il}^{(t)}=1$.
Since the sum utility value (\ref{max1_2}a) is nondecreasing in each iteration according to (\ref{convergenceProofIULP}) and the sum utility value (\ref{max1_2}a) is finitely upper-bounded from (\ref{convergenceProofIULP1_2}), the IULP algorithm must converge.

\section{}
According to \cite{boyd2004convex}, the optimal solution should satisfy the following KKT conditions of problem (\ref{Furmax2}):
\begin{subequations}\label{FurKKT1}
\begin{align}
&\frac{\partial \mathcal L_3}{y_{ij}}=
\frac{\omega_j}{y_{ij}}-\frac{{\omega_j} {q_{ij}g_{ij}}}{{I_{ij}y_{ij}^2}\left(1+\frac{q_{ij}g_{ij}}{I_{ij}y_{ij}}\right) \ln\left(1+\frac{q_{ij}g_{ij}}{I_{ij}y_{ij}}\right)}-\psi_i=0,
\nonumber\\
&\qquad\qquad
\quad\forall j \in \mathcal J_i
\\
&\frac{\partial \mathcal L_3}{q_{ij}}=
\frac{ {\omega_jg_{ij}}}{{I_{ij}y_{ij}}\left(1+\frac{q_{ij}g_{ij}}{I_{ij}y_{ij}}\right) \ln\left(1+\frac{q_{ij}g_{ij}}{I_{ij}y_{ij}}\right)}-\phi_i=0,
\nonumber\\
&\qquad\qquad\quad\forall j \in \mathcal J_i.
\end{align}
\end{subequations}

From (\ref{FurKKT1}a), we obtain
\begin{equation}\label{FurKKT2}
1-\frac{\psi_i y_{ij}}{\omega_j}=\bar f\left(\frac{q_{ij}g_{ij}}{I_{ij}y_{ij}}\right), \quad\forall j \in \mathcal J_i,
\end{equation}
where function $\bar f(x)=\frac{x}{(1+x)\ln(1+x)}$, $x>0$.
Then,
\begin{equation}
\bar f'(x)=\frac{\ln(1+x)-x}{(1+x)^2\ln^2(1+x)}<0, \quad\forall x >0,
\end{equation}
which implies that $\bar f(x)$ is monotonically decreasing.
Since $\lim_{x\rightarrow 0+}\bar f(x)=1$, $\bar f(x)<1$ for any $x>0$.
Obviously, we can obtain that the optimal solution to problem (\ref{Furmax2}) satisfies $y_{ij}>0$ and $q_{ij}>0$, $\forall j \in \mathcal J_i$.
Based on (\ref{FurKKT2}), we must have $\psi_i>0$.
Moreover, according to (\ref{FurKKT1}b), we have
\begin{equation}\label{FurKKT3}
\frac{\phi_i q_{ij}}{\omega_j}=\bar f\left(\frac{q_{ij}g_{ij}}{I_{ij}y_{ij}}\right),\quad \forall j \in \mathcal J_i.
\end{equation}
Since $\bar f(x)=\frac{x}{(1+x)\ln(1+x)}>0$ for any $x>0$, we have $\phi_i>0$ according to (\ref{FurKKT3}).
Based on (\ref{FurKKT2}) and (\ref{FurKKT3}), we have
\begin{equation}\label{FurKKT4}
q_{ij}=\frac{\omega_j-\psi_i y_{ij}}{\phi_i}, \quad\forall j \in \mathcal J_i.
\end{equation}
Plugging (\ref{FurKKT4}) into (\ref{FurKKT2}) yields
\begin{equation}\label{FurKKT4_2}
1-\frac{\psi_i y_{ij}}{\omega_j}-\bar f\left(\frac{ \omega_jg_{ij}}{I_{ij}y_{ij}\phi_i} -\frac{\psi_ig_{ij}}{I_{ij}\phi_i}\right)=0 , \quad\forall j \in \mathcal J_i.
\end{equation}
Define
\begin{equation}\label{FurKKT4_3}
h_{ij}(y_{ij}, \psi_i, \phi_i)=1-\frac{\psi_i y_{ij}}{\omega_j}-\bar f\left(\frac{ \omega_jg_{ij}}{I_{ij}y_{ij}\phi_i} -\frac{\psi_ig_{ij}}{I_{ij}\phi_i}\right).
\end{equation}
Owing to the fact that $\psi_i>0$, $\phi_i>0$ and function $\bar f(x)$ is decreasing, function $h_{ij}(y_{ij}, \psi_i, \phi_i)$ is decreasing in $y_{ij}$ with given $\psi_i$ and $\phi_i$.
Thus, given $\psi_i$ and $\phi_i$, equation (\ref{FurKKT4_2}) is rewritten as
\begin{equation}\label{FurKKT4_3_1}
h_{ij}(y_{ij}, \psi_i, \phi_i)=0,
\end{equation}
which can be uniquely solved by using the bisection method.
Denote the solution to (\ref{FurKKT4_3_1}) by
\begin{equation}\label{FurKKT4_5}
y_{ij}=\hat h_{ij}(\psi_i, \phi_i), \quad\forall j \in \mathcal J_i.
\end{equation}
Substituting (\ref{FurKKT4_5}) into (\ref{FurKKT4}) yields
\begin{equation}\label{FurKKT4_6}
q_{ij}=\frac{\omega_j-\psi_i \hat h_{ij}(\psi_i, \phi_i)}{\phi_i}, \quad\forall j \in \mathcal J_i.
\end{equation}

Finally, the new values of the Lagrange multipliers $\psi_i$ and $\phi_i$ are updated by
\begin{eqnarray}
&&\!\!\!\!\!\!\!\!\!\!\!\!\!\!\!\!\!\!
\psi_i(t+1)=\left[\psi_i(t)+ \kappa(t)\left(\sum_{j\in\mathcal J_i}y_{ij}(t+1)-1\right)\right]^+
,\label{FurKKT10}
\\
&&\!\!\!\!\!\!\!\!\!\!\!\!\!\!\!\!\!\!
\phi_i(t+1)=\left[\phi_i(t)+ \kappa(t)\left(\sum_{i\in\mathcal J_i}q_{ij}(t+1)-p_i^*\right)\right]^+,\label{FurKKT11}
\end{eqnarray}
where $\kappa(t)>0$ is a dynamically chosen stepsize sequence.

\bibliographystyle{IEEEtran}
\bibliography{IEEEabrv,MMM}

\end{document}